\documentclass[12pt]{article}
\usepackage{graphicx}
\usepackage{subfigure}
\usepackage{amsfonts}
\usepackage{amssymb,amsmath}
\usepackage{accents}

\newcommand{\nn}{\nonumber \\}
\newcommand{\bea}{\begin{eqnarray}}
\newcommand{\eea}{\end{eqnarray}}

\makeatletter
\def\widebar{\accentset{{\cc@style\underline{\mskip10mu}}}}
\makeatother

\makeatletter
\@addtoreset{equation}{section}

\makeatother

\usepackage{multicol}
\usepackage[usenames,dvipsnames]{xcolor}
\definecolor{rosso}{cmyk}{0,1,1,0.3}
\definecolor{verde}{cmyk}{0.8,0,0.6,0.25}
\definecolor{bluc}{cmyk}{1,0.4,0,0.1}
\definecolor{blucc}{cmyk}{0.8,0.3,0,0}

\def\be{\begin{equation}}
\def\ee{\end{equation}}

\def\({\left(}
\def\){\right)}
\def\1{^{(1)}}
\def\2{^{(2)}}

\def\<{\langle}
\def\>{\rangle}

\begin{document}

\begin{titlepage}

\begin{flushright}
UT-13-16
\end{flushright}

\vskip 3cm

\begin{center}

{\Large \bf 
Scalar Trapping and Saxion Cosmology
}

\vskip .5in

{
Takeo Moroi$^{(a,b)}$,
Kyohei Mukaida$^{(a)}$,
Kazunori Nakayama$^{(a,b)}$\\
and
Masahiro Takimoto$^{(a)}$
}

\vskip .3in

{\em
$^a$Department of Physics, University of Tokyo, Bunkyo-ku, Tokyo 113-0033, Japan \vspace{0.2cm}\\
$^b$Kavli Institute for the Physics and Mathematics of the Universe,
University of Tokyo, Kashiwa 277-8583, Japan \\
}

\end{center}

\vskip .5in

\begin{abstract}

  We study in detail the dynamics of a scalar field in thermal bath
  with symmetry breaking potential.  In particular, we focus on the
  process of trapping of a scalar field at an enhanced symmetry point
  through the thermal/non-thermal particle production, taking into
  account the interactions of produced particles with the standard
  model particles.  As an explicit example, we revisit the saxion
  dynamics with an initial amplitude much larger than the Peccei-Quinn
  scale and show that the saxion trapping phenomenon happens for the
  most cases and it often leads to thermal inflation.  We also study
  the saxion dynamics after thermal inflation, and it is shown that
  thermal dissipation effect on the saxion can relax the axion
  overproduction problem from the saxion decay.

\end{abstract}

\end{titlepage}

\setcounter{page}{1}

\tableofcontents

%%%%%%%%%%%%%%%%%%%%%%%%%%%%%%%%%%%%%%%%%%%%%%%%%%
\section{Introduction}
\label{sec:intro}
%%%%%%%%%%%%%%%%%%%%%%%%%%%%%%%%%%%%%%%%%%%%%%%%%%

After the discovery of the Higgs boson at the LHC~\cite{Aad:2012tfa},
the concept of spontaneous symmetry breaking induced by the vacuum
expectation value (VEV) of a scalar field becomes more important.
Actually, apart from the electroweak symmetry breaking, the Universe
may have experienced several phase transitions such as the
Peccei-Quinn (PQ) phase transition~\cite{Peccei:1977hh}, grand unified
theory (GUT) phase transition and so on.

Most of these phase transitions are thought to be associated with VEVs
corresponding scalar fields, $\phi$.  While the concept is simple, the
dynamics of the scalar fields leading to the symmetry breaking might
be rather complicated and may have significant cosmological
implications.  The usual argument is as
follows~\cite{Kirzhnits:1976ts}.  At high temperature, $\phi$ sits at
the origin $\phi = 0$ due to thermal effective potential since it is
an enhanced symmetry point.  As the temperature decreases, thermal
effects become less significant, and finally $\phi$ relaxes to the
minimum.  However, the assumption that $\phi$ starts at the point
$\phi=0$ may not always be justified, because $\phi$ can have very
large initial value during and just after inflation.  Then we need to
study dynamics of a scalar in order to see whether or not the symmetry
is restored (i.e., $\phi=0$) in the early Universe.

If the initial amplitude of the scalar field is displaced from the
minimum of the potential, the subsequent dynamics of the scalar field
is non-trivial.  With such an initial condition, the scalar field
oscillates and dissipates its energy via the interaction with
background.  If the effect of the cosmic expansion is negligible, the
scalar field eventually relaxes to an equilibrium state, which is
determined by the initial energy (and other conserved quantities).  In
such a case, the scalar field is expected to be trapped at the
enhanced symmetry point if the temperature of the equilibrium state is
higher than the critical temperature for the phase transition.  On the
contrary, if the cosmic expansion alone efficiently reduces the energy
of oscillating scalar field, the scalar field cannot experience the
restoration of the symmetry and results in the symmetry breaking
vacuum.  The actual situation lies somewhere between these two extreme
cases.  Therefore, to see whether the scalar field experiences the
trapping or not, we have to study in detail the dynamics of scalar
field and compare the time scale of dynamics with that of cosmic
expansion.  On this respect, the phenomenon called non-thermal phase
transition~\cite{Kofman:1995fi} and moduli
trapping~\cite{Kofman:2004yc} was studied previously.

The complete analysis of dynamics of scalar field in a concrete setup
including interactions with standard model (SM) sector in thermal
environment has been lacking.  In a realistic setup, such a scalar
field often interacts with gauge-charged matters and the background is
filled with thermal plasma after inflation.  The primary purpose of
this paper is to analyze the scalar dynamics with symmetry breaking
potential by taking into account all possible relevant effects:
thermal correction to the scalar potential, dissipative effects on the
scalar field, non-perturbative particle production and the resulting
modification on the scalar potential.

To make our discussion concrete, we will pay particular attention to
the dynamics of saxion in a class of supersymmetric (SUSY) axion
model.  We will show that, even if the initial amplitude of the saxion
is much larger than the PQ scale, the zero-mode of the saxion field is
eventually trapped at the symmetry enhanced point in large parameter
space; this is due to the modification of the saxion potential and
the dissipation of the oscillation.  Thus, the thermal
inflation~\cite{Yamamoto:1985rd,Lyth:1995hj,Lyth:1995ka} is quite
often caused by the saxion.  We also study the scalar dynamics after
phase transition.  We find the situation where thermal dissipation
into SM plasma efficiently reduces the saxion energy density without
axion overproduction.

This paper is organized as follows.  In Sec.~\ref{sec:pre}, we
introduce a basic setup and review thermal effects and particle
production processes as preparations for the analyses in the following
sections.  In Sec.~\ref{sec:saxion}, we study the saxion dynamics in
detail in the cases of both small and large mixing between PQ and SM
quarks.  We will study in detail whether the saxion is trapped at the
origin or not by taking account for thermal and non-thermal effects in
the following sections (Sec.~\ref{sec:small} and
Sec.~\ref{sec:large}).  In Sec.~\ref{sec:PT}, we will discuss the
saxion dynamics after thermal inflation and its cosmological
consequence.  Sec.~\ref{sec:conc} is devoted to the conclusion and
discussion.

%%%%%%%%%%%%%%%%%%%%%%%%%%%%%%%%%%%%%%%%%%%%%%%%%
\section{Preliminaries}
\label{sec:pre}
%%%%%%%%%%%%%%%%%%%%%%%%%%%%%%%%%%%%%%%%%%%%%%%%%

%%%%%%%%%%%%%%%%%%%%%%%%%%%%%%%%%%%%%%%%%%%%%%%%%
\subsection{Overview}
%%%%%%%%%%%%%%%%%%%%%%%%%%%%%%%%%%%%%%%%%%%%%%%%%

Here, we consider the evolution of a complex scalar field which is
responsible for the spontaneous breaking of a global $U(1)$ symmetry;
the (zero-temperature) potential of $\phi$ has a minimum at
$|\phi|=v\neq 0$.  Above this scale, the potential of $\phi$, which is
lifted by the SUSY breaking effect, is assumed to be quadratic:
\begin{align}
  V \simeq
  m^2 |\phi|^2 ~~\mbox{for}~~|\phi| \gg v.
\end{align}
In addition, the scalar $\phi$ is assumed to have interactions with
scalar $\tilde Q$ and fermion $Q$ as
\begin{align}
  \mathcal L =  \lambda_s^2|\phi|^2|\tilde Q|^2 +  
  (\lambda_f \phi {\widebar{Q}} Q + {\rm  h.c.}), 
\end{align}
with coupling constants $\lambda_s$ and $\lambda_f$.  $\tilde Q$ and
$Q$ are assumed to have SM gauge interactions.  Having a SUSY model in
mind, we call $Q$ and $\tilde Q$ as quark and squark, respectively,
hereafter.

How does the phase transition proceed?  As described in the
Introduction, the usual arguments are as follows.  At high
temperature, $\phi$ sits at the origin $\phi = 0$ due to thermal
effective potential.  In our case, it may be caused by the interaction
with $\tilde Q$ and $Q$.  As the temperature decreases, thermal
effects become less significant, and finally $\phi$ relaxes to the
minimum $|\phi|=v$.

However, it is non-trivial whether $\phi$ actually starts at the point
$\phi=0$.  Let us assume that initially, just after inflation, $\phi$
is placed far from the origin: $|\phi| \gg v$.  This assumption is
quite natural if $m \ll H_{\rm inf}$, where $H_{\rm inf}$ denotes the
Hubble scale during inflation.  An immediate consequence is that since
$\tilde Q$ and $Q$ get large masses from the VEV of $\phi$, they may
decouple from thermal bath even after the reheating, depending on the
reheating temperature.  Thus we need to investigate the dynamics of
$\phi$ in order to see whether $\phi$ will be eventually trapped at
$\phi=0$ or not.
\begin{itemize}
\item If $|\phi|$ is large enough, $\tilde Q$ and $Q$ decouple from
  thermal bath.  Still, however, $\phi$ feels thermal effects through
  the so-called thermal logarithmic potential~\cite{Anisimov:2000wx},
  which may tend to stabilize $\phi$ toward the origin.
\item The dissipation is essential to effectively reduce the coherent
  oscillation energy density~\cite{Berera:1995ie+X,Anisimov:2000wx, Yokoyama:2005dv,Drewes:2010pf,BasteroGil:2010pb,Moroi:2012vu,
    Mukaida:2012qn}.  In particular, we must take into account thermal
  dissipation effects on the scalar in order to see $\phi$ truly
  relaxes to the origin.
\item In the limit of low reheating temperature, such thermal effects
  would be inefficient.  Even in such a case, because of the large
  amplitude of $\phi$, non-perturbative particle production events
  happen at each oscillation of $\phi$.  Then, the finite number
  density of $\tilde Q$ and $Q$ modifies the effective potential of
  $\phi$ which also tends to stabilize $\phi$ at the origin.
\item The produced $\tilde Q$ and $Q$ decay or annihilate depending on
  their interactions to the SM particles.  If $Q$ has the
  same gauge quantum numbers as of right-handed down-type quarks, for
  example, we may obtain the following mixing term
  \begin{equation}
    \mathcal L_{\rm mix} = \kappa Q q_L H
    + {\rm h.c.},
    \label{L_mix}
  \end{equation}
  where $q_L$ and $H$ are SM left-handed quark and Higgs boson,
  respectively.  With the above interaction, $Q$ soon decays and
  generates thermal plasma.  Even without such mixings, annihilation
  processes into SM particle also reduce $\tilde Q$ and $Q$, which
  would also generate thermal plasma.  Through these processes, $\phi$
  coherent oscillation effectively lose its energy by producing
  thermal plasma, which may significantly affect the scalar dynamics.
\end{itemize}

Thus the actual dynamics of the scalar $\phi$ would be complicated due
to the combinations of these effects.  We will analyze the scalar
dynamics in Sec.~\ref{sec:saxion} with an explicit example of the
saxion in SUSY axion model, but we emphasize that the analysis can be
applied to general class of scalar fields with symmetry breaking
potential.  In the rest of this section, we briefly summarize the
thermal effects and particle production processes.

%%%%%%%%%%%%%%%%%%%%%%%%%%%%%%%%%%%%%%%%%%%%%%%%%
\subsection{Thermal and non-perturbative effects on scalar dynamics}
\label{sec:ds}
%%%%%%%%%%%%%%%%%%%%%%%%%%%%%%%%%%%%%%%%%%%%%%%%%

When the primordial inflation ends, the scalar field $\phi$ may be
displaced far from the true vacuum $|\phi| = v$.  Eventually, it
begins to relax towards the equilibrium state which is not necessarily
identical to the true vacuum because there is a background thermal
plasma.  Even with this initial condition, which is far from the
origin of potential, the scalar field may be trapped at the origin due
to thermal effects since the (s)quarks become massless at the origin
of scalar potential.  In addition, even if thermal effects are not
efficient, the scalar field can be trapped by the explosive production
of squarks via the parametric
resonance~\cite{Kofman:1994rk,Kofman:1997yn}.

Now let us summarize the effect of background thermal plasma at the
beginning of scalar coherent oscillation and also the conditions for
non-perturbative particle production in the presence of thermal
plasma~\cite{Mukaida:2012qn,Enqvist:2012tc}.  We will discuss whether
the saxion is trapped at the origin or not by taking account for
thermal and non-thermal effects in the following sections
(Sec.~\ref{sec:small} and Sec.~\ref{sec:large}).

%%%%%%%%%%%%%%%%%%%%%%%%%%%%%%%%%%%%%%%%%%%%%%%%%
\subsubsection{Thermal effects at the beginning of oscillation}
\label{sec:th_begin}
%%%%%%%%%%%%%%%%%%%%%%%%%%%%%%%%%%%%%%%%%%%%%%%%%

Since the (s)quarks are in the thermal plasma if $\lambda |\phi|
\lesssim T$, there might sink at the origin of free energy.  At the
one-loop level, the free energy is given by \cite{Dolan:1973qd}
\begin{align}
  V_{\rm 1-loop} = \sum_{i} (-)^{|i|}
  \frac{T^4}{\pi^2} \int_0^\infty dz\, z^2
  \ln \left[ 1 - (-)^{|i|} e^{- \sqrt{z^2 + M_i^2 (|\phi|) / T^2}} \right]
\end{align}
where $i$ represents the species (normalized by one complex scalar or
one chiral fermion), $|i|$ denotes the statistical property: $|i| =
0,1$ for boson and fermion respectively, and $M_i$ is the mass of $i$
that depends on $|\phi|$.  If the mass is smaller than the temperature
$M_i \ll T$, then one can find that this term encodes the so-called
``{\it thermal mass}'' term:
\begin{align}
	V_{\rm 1-loop} \supset 
	\sum_{i \in {\rm Boson}} \frac{1}{12} T^2 M_i^2 (|\phi|)
	+ \sum_{i \in {\rm Fermion}} \frac{1}{24} T^2 M_i^2 (|\phi|).
\end{align}

On the other hand, if $M_i \gg T$, this one-loop contribution rapidly
vanishes. Instead, higher loop effects dominate the free energy, which
arise from the threshold correction to the gauge coupling constant $g$
at $T$.  That is, using the fact that the free energy of hot plasma
has a contribution proportional to $g^2 (T) T^4$, and that the gauge
coupling constant at the scale below $M_i$ has a logarithmic
dependence on the scale $M_i$, one finds the free energy of $\phi$,
so-called ``{\it thermal log}'' potential as~\cite{Anisimov:2000wx}:
\begin{align}
  V_{\rm th-log} = a_{\rm L}\, \alpha (T)^2 T^4 
  \ln \left[ \lambda^2 |\phi|^2 /T^2 \right]
\end{align}
where $a_{\rm L}$ is an order one constant.

Thus, the free energy that affects the dynamics of scalar condensation can be approximately 
parametrized as
\begin{align}
	V_{\rm th} = \begin{cases}
	a_{\rm M}\, \lambda^2 T^2 |\phi|^2 &\mbox{for}~~\lambda |\phi| < T \\[5pt]
	a_{\rm L}\, \alpha (T)^2 T^4 \ln \left[ \lambda^2 |\phi|^2 /T^2 \right] 
	&\mbox{for}~~\lambda |\phi| > T
	\end{cases}
\end{align}
where $a_{\rm M/L}$ are order one constants.
The effective potential for the scalar condensation is given by the sum: 
$V_{\rm eff} = V + V_{\rm th}$.

The scalar condensation begins to oscillate when the Hubble parameter
becomes comparable to its effective mass:
\begin{align}
	H_{\rm os} \simeq \max
	\left[ 
		m,
		\lambda T~(\mbox{for}~\lambda \phi_i < T),
		\alpha T^2/\phi_i~(\mbox{for}~T < \lambda \phi_i)
	\right]
\end{align}
with the initial amplitude $\phi_i$ being assumed to be larger than $v$.
Here we omit the order one constants $a_{\rm M/L}$ for simplicity.
There are three cases depending on which term dominates the effective potential:
thermal log, thermal mass and zero temperature mass.
The temperature $T_{\rm os}$ at the beginning of oscillation
for each cases is summarized as follows~\cite{Mukaida:2012qn}:
\begin{itemize}
\item The scalar $\phi$ begins to oscillate with thermal log potential
  if
	\begin{align}
		\phi_i < \alpha T_{\rm R} \sqrt{\frac{M_{\rm pl}}{m}}
		~~~\mbox{and}~~~
		\phi_i > \(T_{\rm R} / \lambda\)^{2/3} \(\alpha M_{\rm pl}\)^{1/3}. \label{eq:bdry_a_b}
	\end{align}
	The temperature at the beginning of oscillation is given by
	\begin{align}
		T_{\rm os} = \sqrt{\frac{\alpha M_{\rm pl}}{\phi_i}} T_{\rm R}.
	\end{align}
	In addition to above two inequalities, the condition $T_{\rm os} > T_{\rm R}$ should be met
	since otherwise the scalar oscillates with the zero-temperature mass term [case (iii)].
	\item The scalar $\phi$ begins to oscillate with thermal mass if
	\begin{align}
	\begin{cases}
		\lambda > \left[ \cfrac{m^3}{T_{\rm R}^2 M_{\rm pl}} \right]^{1/4}
		~~\mbox{and}~~~
		\phi_i < \(T_{\rm R} / \lambda\)^{2/3} \(\alpha M_{\rm pl}\)^{1/3}
		&\mbox{for}~~\lambda M_{\rm pl} > T_{\rm R} \\[20pt]
		\lambda > \sqrt{\cfrac{m}{M_{\rm pl}}}
		&\mbox{for}~~\lambda M_{\rm pl} < T_{\rm R} .
		\end{cases}
	\end{align}
	The temperature at the onset of oscillation is given by
	\begin{align}
		T_{\rm os} = 
		\begin{cases}
			\left[ \lambda T_{\rm R}^2 M_{\rm pl} \right]^{1/3} 
			&\mbox{for}~~\lambda M_{\rm pl} > T_{\rm R} \\[5pt]
			\lambda M_{\rm pl}
			&\mbox{for}~~\lambda M_{\rm pl} < T_{\rm R}.
		\end{cases}
	\end{align}
	\item
	Otherwise, the scalar $\phi$ begins to oscillate with zero temperature mass,
	and the temperature is given by
	\begin{align}
		T_{\rm os} =
		\begin{cases}
			\left[ m M_{\rm pl} T_{\rm R}^2  \right]^{1/4}
			&\mbox{for}~~m M_{\rm pl} > T_{\rm R}^2\\[5pt]
			\sqrt{m M_{\rm pl}}
			&\mbox{for}~~m M_{\rm pl} < T_{\rm R}^2.
		\end{cases}
	\end{align}
\end{itemize}
%%

%%%%%%%%%%%%%%%%%%%%%%%%%%%%%%%%%%%%%%%%%%%%%%%%%
\subsubsection{Non-perturbative particle production}
\label{sec:NP}
%%%%%%%%%%%%%%%%%%%%%%%%%%%%%%%%%%%%%%%%%%%%%%%%%

The dispersion relations of (s)quarks depend on the scalar field value:
\begin{align}
  \omega_Q (t) = \sqrt{\lambda^2 |\phi (t)|^2 + m_{{\rm eff}, Q}^2 + {\bf k}^2}
\end{align}
with the soft (s)quark mass being neglected.  Here $m_{{\rm eff},Q}$
encodes the finite density correction to the real part of dispersion
relation: {\it e.g.} the contribution from particles in the thermal
plasma is imprinted as the thermal mass term.  In our case, the
initial amplitude $\phi_i$ is much larger than the scalar mass $m$:
$\lambda \phi_i \gg m$, and hence perturbative decay is kinematically
forbidden during the most region of oscillation period.  Then, the
following non-perturbative particle production becomes important.

The non-perturbative particle production occurs when the adiabaticity
of (s)quarks is broken down: $|\dot{\omega}_Q/\omega_Q^2| \gg
1$~\cite{Kofman:1994rk,Kofman:1997yn}.  The condition for
non-perturbative production is given by
\begin{align}
  \lambda \tilde \phi \gg \max \left[ m, \frac{m_{{\rm eff},Q}^2}{m} \right],
  \label{eq:np_cond}
\end{align}
where $\tilde \phi$ is the amplitude of the oscillation of $\phi$.
Note that even if the back-reaction of particle production dominates
the dynamics of scalar, the condition [Eq.~\eqref{eq:np_cond}] is
useful by simply replacing $m$ with the effective mass $\bar m_{\rm
  eff}$ that encodes the finite density correction.

If Eq.~\eqref{eq:np_cond} is met, after the first crossing of the
region where the adiabaticity is broken: $|\phi| < \phi_{\rm NP}$ with
$\phi_{\rm NP}\equiv (m \tilde \phi /\lambda)^{1/2}$, the number
density of (s)quark suddenly acquires the value
\begin{align}
  n_{Q} \simeq \frac{k_\ast^3}{4 \pi^3};~~~k_\ast\equiv (\lambda m \tilde \phi)^{1/2}
\end{align}
for one complex scalar or one chiral fermion~\cite{Kofman:1997yn}.  Before the
first crossing of the region where the adiabaticity is broken, we have
$n_Q = 0$.  Thus, from Eq.\eqref{eq:np_cond}, we obtain the condition
for the non-perturbative production:
\begin{align}
	k_\ast^2 \gg \max \left[ m^2, g^2 T^2 \right].
	\label{NP:begin}
\end{align}
In parameters of our interest,\footnote{
	Here we assumed $\lambda \lesssim g^2$.
	See also footnote \ref{ft:lam_alp}.
}
the non-perturbative production may occur when the scalar $\phi$
begins to oscillate with the zero-temperature mass,
and the reheating takes place after the beginning of oscillation.
Hence, at the first crossing, the latter condition of Eq.~\eqref{NP:begin} implies 
\begin{equation}
	T_{\rm R} < (\lambda / g^2 ) \sqrt{\frac{m}{M_{\rm pl}}} \phi_i. \label{eq:bdry_b_c}
\end{equation}
Then there are three cases. %[See Fig.~\ref{fig:Tos}.]
\begin{itemize}
\item[(A)] $\phi$ begins to oscillate with thermal log potential.
\item[(B)] $\phi$ begins to oscillate with zero-temperature potential
  without non-perturbative particle production.
\item[(C)] $\phi$ begins to oscillate with zero-temperature potential
  with efficient non-perturbative particle production.
\end{itemize}
We discuss the scalar trapping in these three cases separately.

%%%%%%%%%%%%%%%
%\begin{figure}[h]
\begin{figure}[t]
\begin{center}
%\vskip -1.cm
\includegraphics[scale=1.5]{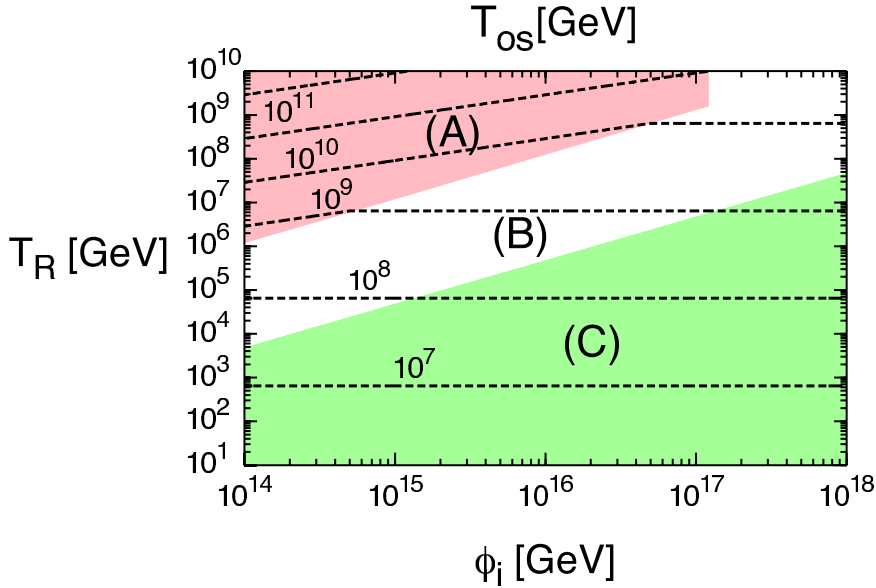}
\vskip -0.1cm
\caption{\small The contour plot of $T_{\rm os}$ as a function of
  $\phi_i$ and $T_{\rm R}$.  Here, we have taken $m=1\ {\rm GeV}$,
  $\lambda=0.05$, and used $\alpha$ of the strong interaction.  In the
  region (A), the scalar begins to oscillate with thermal log
  potential.  In the region (B), the scalar begins to oscillate with
  zero-temperature mass without non-perturbative particle production.
  In the region (C), the scalar begins to oscillate with
  zero-temperature mass with efficient non-perturbative particle
  production.  }
\label{fig:Tos}
\end{center}
\end{figure}
%%%%%%%%%%%%%%%

In Fig.~\ref{fig:Tos}, we show the contours of constant $T_{\rm os}$
on $\phi_i$ vs.\ $T_{\rm R}$ plane; here, we used $m=1\ {\rm GeV}$,
$\lambda=0.05$, and $\alpha$ of the strong interaction.  The shaded
region (A) corresponds to the parameters where the scalar begins to
oscillate with the thermal log term.  On the other hand, in the
regions (B) and (C), the scalar begins to oscillate with the zero
temperature mass term.  The boundary between (A) and (B) [(B) and (C)]
is given by Eq.\eqref{eq:bdry_a_b} [Eq.~\eqref{eq:bdry_b_c}].  Note
that the region where the scalar starts to oscillate with the thermal
mass term lies outside of this figure: higher $T_{\rm R}$ and smaller
$\phi_i$.  We will not consider this case in the following.

The subsequent evolution of scalar/plasma system after the first
crossing crucially depends on the setup.  There are two cases:
\begin{enumerate}
\item If the produced particles cannot decay faster than the
  oscillation period, the parametric resonance can occur due to the
  induced emission effect of previously produced particles.  Hence,
  the number density grows exponentially and then the back-reaction
  soon dominates the effective mass of $\phi$.  In addition, because
  of the largeness of the number density of $Q$ in the background, the
  linear potential of $\phi$ is induced. [See Eq.~\eqref{eq:linear_1}.]
  This case will be extensively studied in Sec.~\ref{sec:small}.
\item On the other hand, if the non-perturbatively produced (s)quarks
  can decay into other light particles much faster than the
  oscillation period of $\phi$, then the phenomena like instant
  preheating~\cite{Felder:1998vq} take place. In this case, the linear
  potential generated by the produced
  particles is insignificant since the produced particles soon
  decay. See Sec.~\ref{sec:large} for further discussion.
\end{enumerate}

First, let us consider the case (i).  After the passage of
non-adiabatic region $|\phi| < \phi_{\rm NP}$, the produced particles
become heavy due to large field value $|\phi (t)|$ and their decay
rate become large correspondingly.  If the decay rate
$\Gamma_{\rm dec}^Q\sim \kappa^2 \lambda |\phi (t)|$ [with $\kappa$
being the mixing parameter given in Eq.\ \eqref{L_mix}] reaches
$\Gamma^Q_{\rm dec} (t_{\rm dec})t_{\rm dec} \sim 1$ with $t_{\rm dec}
\ll m^{-1}$, the produced particles completely decay into SM radiation
well before the scalar moves back to the origin.  
The condition for the case (i), which requires the stability of $Q$
for the time scale of our interest, is given by
\begin{align}
  \kappa \ll \sqrt{\frac{m}{\lambda \tilde \phi}}.
  \label{eq:notdecay}
\end{align}
If the above condition is satisfied, the number density of (s)quarks
grows explosively due to the parametric resonance as $\phi$ crosses
the region where the adiabaticity is broken.  In our set up, the
condition Eq.~\eqref{eq:np_cond} implies
\begin{align}
  k_\ast^2 \gg \max
  \left[\bar m_{\rm eff}^2,g^2 T^2, g^2 \frac{n_Q}{k_\ast},
    \lambda^2 \frac{n_Q}{k_\ast}  \right]
\end{align}
where $\bar m_{\rm eff}$ represents the effective mass for $\phi$ that
includes the back-reaction of particle creation and it is estimated as
\begin{align}
  \bar m^2_{\rm eff} \sim \max \left[ m^2, 
    \lambda \frac{n_Q}{\tilde \phi} \right]. \label{eq:eff_mass}
\end{align}
Eventually, the parametric resonance stage finishes when the
inequality Eq.~\eqref{eq:np_cond} is saturated.  This condition can be
rewritten as
\begin{align}
  \bar m_{\rm eff} \ll \min \left[ k_\ast, \frac{\lambda^2}{g^2} k_\ast \right] 
  \equiv  \frac{k_*}{c}.
  \label{eq:pr_finish}
\end{align}
In the following, we consider the case: $\lambda \lesssim g^2$ for simplicity,\footnote{
  For large $\lambda$, 
  $\phi$ may be more likely to be trapped at the origin
  since the region (C) where the non-perturbative production occurs
  becomes larger [See Eq.~\eqref{eq:bdry_b_c}].
  \label{ft:lam_alp}
}and
hence $c = g^2/\lambda^2$. In addition, in the case (i), the linear potential of $\phi$, which is 
obtained as~\cite{Kofman:1997yn}
\begin{align}
  V_{\rm linear} \sim
  \lambda^2 |\phi|^2 \langle |\tilde{Q}|^2 \rangle
  \sim
  \lambda^2 |\phi|^2 \int \frac{d^3 k}{(2\pi)^3\omega_Q} f _Q(k) \sim
  \lambda |\phi| n_Q~~~
  \mbox{for}~~|\phi| > \phi_{\rm NP}, \label{eq:linear_1}
\end{align}
is generated because sizable amount of $Q$ exists irrespective of the
field value of $\phi$.\footnote
{Here we used the fact that the $k$ integration is dominated at $k\sim
  k_*$ and $\omega_Q\sim \lambda|\phi|$.  For the case of
  $|\phi|\lesssim\phi_{\rm NP}$, $\omega_Q\sim k_*$, and hence the
  potential becomes quadratic in $|\phi|^2$.}
(Here, $f_Q$ denotes the distribution function of $Q$.)  Such a linear
potential plays very important role in studying the evolution of
$\phi$, as we will see in the next section.

Next, let us consider the case (ii), which corresponds to the case
where the condition \eqref{eq:notdecay} is not satisfied.  In this
case, the produced particles soon decay at
$|\phi (t_{\rm dec})| \sim [m \tilde \phi/ (\kappa^2\lambda)]^{1/2}
\ll \tilde \phi$.  Taking an oscillation time average, one finds the
effective dissipation rate in this case~\cite{Mukaida:2012qn}:
\begin{align}
  \Gamma_\phi^{\rm (dis)} \simeq N_{\rm d.o.f.} \times 
  \frac{\lambda^2 m}{ 2 \pi^4 \kappa}
\end{align}
where $N_{\rm d.o.f.}$ stands for the degree of freedom normalized by
one complex scalar or one chiral fermion.  In addition, in case (ii),
the parametric resonance does not occur and the linear potential
generated by the produced particles is insignificant.

%%%%%%%%%%%%%%%%%%%%%%%%%%%%%%%%%%%%%%%%%%%%%%%%%%
\section{Saxion Dynamics}
\label{sec:saxion}
%%%%%%%%%%%%%%%%%%%%%%%%%%%%%%%%%%%%%%%%%%%%%%%%%%

The strong CP problem is one of the serious theoretical problems in
the SM.  The PQ mechanism is a promising solution to the strong CP
problem~\cite{Peccei:1977hh}.  It utilizes a global U(1) symmetry,
called PQ symmetry, which is spontaneously broken at the scale $f_a$.
The axion appears as a consequence of spontaneous breakdown of the PQ
symmetry and it plays an important role in cosmology and
phenomenology~\cite{Kim:1986ax,Kawasaki:2013ae}.

In SUSY extensions of the axion model, the early Universe cosmology
becomes much more non-trivial because of the presence of flat
direction in the scalar potential, called
saxion~\cite{Kugo:1983ma,Tamvakis:1982mw,Kim:1984yn,Rajagopal:1990yx}.
The cosmological evolution of the saxion field depends on the detailed
structure of the saxion potential.  Among various possibilities (see
Ref.~\cite{Kawasaki:2013ae} for a partial list of the stabilization
mechanism), in this paper we focus on the case where there is only one
(complex) PQ scalar field having VEV
\cite{Asaka:1998ns,Abe:2001cg,Nakamura:2008ey,Jeong:2011xu}.

The potential of the saxion field $\phi$ is lifted only by the effect
of SUSY breaking and is very flat for $|\phi| \gg f_a$.  Then, without
a tuning, the saxion is expected to be initially placed at the origin
$(|\phi| =0)$ or around the Planck scale $(|\phi|\sim M_{\rm pl})$,
depending on whether the saxion receives positive or negative Hubble
mass correction during inflation.  In the former case, the saxion
likely causes thermal
inflation~\cite{Yamamoto:1985rd,Lyth:1995hj,Lyth:1995ka} as
extensively studied in
Refs.~\cite{Choi:1996vz,Chun:2000jr,Kim:2008yu,Choi:2009qd,Park:2010qd}.
Once thermal inflation happens, the saxion must successfully reheat
the Universe before the big-bang nucleosynthesis (BBN) begins.
Unfortunately, the saxion often dominantly decays into the axion pair,
which would invalidate the thermal inflation scenario.  (However, it
is possible that the saxion dominantly decays into the Higgs sector in
the DFSZ model~\cite{Dine:1981rt}.)

On the other hand, the case of large initial amplitude of the saxion,
$|\phi| \gg f_a$, has not been studied in detail so far.  We mainly
focus on this case and investigate its cosmological effects.  The
dynamics of the saxion in such a case is rather complicated due to two
effects: thermal effects and non-perturbative particle
production.\footnote{ Thermal effects on the saxion dynamics were
  considered in Refs.~\cite{Kawasaki:2010gv,Moroi:2012vu} in a
  different class of PQ model.  } As will become clear, both effects
tend to temporary stabilize the saxion around the origin, $|\phi|=0$.
We will show that, even if we start with a large saxion initial
amplitude, it will be eventually trapped at the origin due to these
effects and thermal inflation takes place for the most reasonable
parameter choices.  Cosmological issues related to this will be
discussed in Sec.\ \ref{sec:PT}.

%%%%%%%%%%%%%%%%%%%%%%%%%%%%%%%%%%%%%%%%%%%%%%%%%%
\subsection{Setup}
\label{sec:setup}
%%%%%%%%%%%%%%%%%%%%%%%%%%%%%%%%%%%%%%%%%%%%%%%%%%

Amongst several ways to stabilize the saxion, we focus on the models
with only one PQ scalar whose potential is stabilized by the quadratic
term from SUSY
breaking~\cite{Asaka:1998ns,Abe:2001cg,Nakamura:2008ey,Jeong:2011xu}.

To capture the essential features of saxion dynamics within these
models, we approximate the saxion potential as
\begin{align}
  V \sim
  \begin{cases}
    - m_{\rm s}^2 \left( |\phi|^2 - M^2 \right) & \mbox{for}~~|\phi| < M \\
    -m_{\rm s}^2M^2 \log^{3} (|\phi|/M) & \mbox{for}~~M< |\phi| < f_a \\
    + m^2 \left( |\phi|^2 - f_a^2 \right) & \mbox{for}~~f_a < |\phi|. \\
  \end{cases}
\end{align}
where $f_a$ is the PQ scale, $m_{\rm s}$ and $m$ are mass parameters
terms from SUSY breaking effect.  (Note that we have the relation
$mf_a \simeq m_{\rm s}M$.)  In addition, $M$ is the messenger scale in
the framework of gauge-mediation.  (For the cases of gravity- and
anomaly-mediation, see the discussion below.)  The relation between
$m_{\rm s}$ and $m$ depends on the mechanism of SUSY breaking.  In the
models based on gauge-mediation~\cite{Asaka:1998ns,
  ArkaniHamed:1998kj, Choi:2011rs, Nakayama:2012zc}, $m$ is expected
to be of order the gravitino mass.  Furthermore, the negative soft
mass of saxion potential at its origin comes from the three loop
effect via the Yukawa coupling $\lambda$, and hence it is related with
the squark mass as $m_{\rm soft} \propto m_{\rm s}/\lambda$ (with
$m_{\rm soft}$ being the sfermion masses in SUSY SM.)  As a result,
there can be a hierarchy between $M$ and $f_a$.  In the gravity- or
anomaly-mediation models, the renormalization group effect may
stabilize the saxion potential~\cite{Abe:2001cg,Nakamura:2008ey}. If
so, $m\simeq m_{\rm s}$ (and $M\simeq f_a$).

\begin{table}[t]
\caption{Charge assignments}
\begin{center}
{\renewcommand\arraystretch{1.8}
  \begin{tabular}{ccccc}
    \hline
    \hline
    &&$\hat \phi$ & $ \hat Q$ & $\hat {\widebar{Q}}$ \\
    \hline
    U(1)$_{\rm PQ}$ && $+1$ & $-1$ & $0$ \\
    SU(3)$_{\rm c}$ && $1$ &  $3$ & $\bar 3$ \\
    \hline
    \hline
  \end{tabular}
}
\label{tab:q_asign}
\end{center}
\label{default}
\end{table}

We consider the SUSY extension of the hadronic axion
model~\cite{Kim:1979if}.  We introduce the following superpotential:
\begin{align}
  W = \lambda \hat \phi \hat{\widebar{Q}} \hat Q,  \label{PQQ}
\end{align}
where $\hat Q$ and $\hat{\widebar{Q}}$ are identified as the PQ quark
superfields\footnote{ Symbols with hats represent the superfield. The
  corresponding scalar component is represented by symbols without
  hats.  } charged under both SU(3)$_{\rm c}$ and U(1)$_{\rm PQ}$ so
that it has U(1)$_{\rm PQ}$--SU(3)$_{\rm c}$--SU(3)$_{\rm c}$ anomaly
and solves the strong CP problem.  The charge assignments are
summarized in Table~\ref{tab:q_asign}.\footnote
{There can also be PQ leptons, $\hat{\widebar{L}}$ and $\hat L$, which
  are combined with PQ quarks to form SU(5) fundamental and
  anti-fundamental representations as $(\hat Q, \hat{\widebar{L}})$
  and $(\hat{\widebar{Q}},\hat L)$.  Even with those particles, the
  following arguments are unchanged.}
(Here, the PQ charges of $\hat Q$ and $\hat{\widebar{Q}}$ are arranged
so that the mixings with the SM fermions are possible.)

With the PQ charges given in Table~\ref{tab:q_asign}, there may exist
the following terms in the superpotential
\begin{equation}
  W_{\rm mix} = \kappa  \hat{\widebar Q} \hat{q}_L \hat{H}_d, \label{eq:mix}
\end{equation}
which induce the mixing with the SM particles.  (Here, all the
particles in the SUSY SM are vanishing PQ charge.)  If these mixing
terms exist, PQ quarks decay into SM fields.  As will become clear,
the dynamics of saxion significantly depends on whether PQ (s)quarks
are stable or not.  In this paper we consider two cases separately:
small mixing limit (Sec.~\ref{sec:small}) and large mixing limit
(Sec.~\ref{sec:large}).

%%%%%%%%%%%%%%%%%%%%%%%%%%%%%%%%%%%%%%%%%%%%%%%%%
\subsection{Saxion dynamics: small mixing case}
\label{sec:small}
%%%%%%%%%%%%%%%%%%%%%%%%%%%%%%%%%%%%%%%%%%%%%%%%%

As mentioned at the beginning of Sec.~\ref{sec:ds},
the saxion may be trapped at the origin due to the thermal/non-thermal effects
even though its initial field value is displaced far from the origin.
Let us discuss thermal and non-thermal effects in turn
and derive the condition for saxion trapping.
In this subsection we focus on the case of small mixing limit between PQ and SM quarks,
and hence PQ (s)quarks are considered as stable particles.

%%%%%%%%%%%%%%%%%%%%%%%%%%%%%%%%%%%%%%%%%%%%%%%%%
\subsubsection{Case (A): Thermal trapping}
\label{sec:trap_th}
%%%%%%%%%%%%%%%%%%%%%%%%%%%%%%%%%%%%%%%%%%%%%%%%%

First, we consider the case (A).  If the negative mass of saxion
potential at the origin vanishes due to the thermal effects when the
Hubble parameter becomes comparable to the dissipation rate of saxion
into radiation, the saxion is trapped at the origin.  In our setup,
the dissipation rate is expected to become larger than the Hubble
parameter soon after the onset of oscillation because the temperature
$T_{\rm os}$ is much larger than $m$ and the coupling $\lambda$ is not
so small.  Hence, we derive the condition for saxion trapping with the
approximation that the saxion immediately dissipates its energy after
its onset of oscillation.  Note that there is no explosive particle
production due to the non-perturbative effect for $\lambda < \alpha$.

Let us consider the first crossing of $\phi=0$.  The time scale
$\delta t$, during which the saxion passes through $|\phi| \lesssim
\phi_c \equiv T_{\rm os}/\lambda$, is estimated as $\delta t \sim
1/(\lambda \alpha T_{\rm os})$.  The abundance of PQ (s)quarks that is
produced during this time interval is given by $n_Q \sim n_{\rm th}^2
\langle\sigma_{\rm prod} v\rangle \delta t$, where $\langle\sigma_{\rm
  prod} v\rangle \sim g^4 /T_{\rm os}^2$ denotes the PQ (s)quark pair
production cross section and $n_{\rm th}\sim T_{\rm os}^3$ is the
number density of particles in thermal bath. Thus the number density
after the first passage for $|\phi| < \phi_c$ is evaluated as
\begin{equation}
  n_Q^{s} \simeq 
  T_{\rm os}^3,
\end{equation}
as long as $\lambda < \alpha$.  (Here, the superscript ``$s$'' is for
the saturated value.)  After the passage, the saxion may climb
  up the linear potential $V_{\rm eff}\simeq \lambda T_{\rm os}^3 |\phi|$
  for $|\phi| > \phi_c$, and eventually reach its maximum value in
  $\delta t_{\rm max} \sim \alpha/ (\lambda T)$.  After the decoupling
  of PQ squarks at $|\phi| > \phi_c$, the annihilation of PQ squarks
  may still be efficient.  In this case, their abundance reduces to
  $n_Q \sim ( \langle\sigma_{\rm ann} v\rangle \delta t_{\rm max}
  )^{-1}$.  The result for $|\phi| > \phi_c$ is
\begin{align}
  n_Q^{l}\simeq {\rm min}\left[1, \frac{\lambda}{\alpha^3}\right] 
  \times T_{\rm os}^3 \equiv d_\lambda T_{\rm os}^3.
\end{align}

Recalling that we are considering the case where PQ (s)quarks are
(quasi-)stable, we obtain the effective potential of the saxion:\footnote
{$V_{\rm eff}$ can be evaluated as $V_{\rm linear}$ given in Eq.\
  \eqref{eq:linear_1}.  Note that, in the present case, the $k$
  integration is dominated at $k\sim T$.}
\begin{align}
  V_{\rm eff} \simeq
  \begin{cases}
    \displaystyle \frac{\lambda n_Q^{s}}{\phi_c} |\phi|^2 & \mbox{for}~~|\phi| < \phi_c \\[10pt]
    \lambda  n_Q^{l} |\phi| & \mbox{for}~~|\phi| > \phi_c.
  \end{cases}
  \label{Veff}
\end{align}
Then, the conservative condition for saxion trapping is summarized as
follows:\footnote
{Throughout this paper, we impose the following conservative
  condition; we require that, for the trapping at the origin, the
  origin be the absolute minimum of the thermal potential.  We do not
  consider the case where there are other minima than the origin.  As
  we will see, even with this conservative criteria, the saxion is
  likely trapped for most parameters of our interest.  }
\begin{itemize}
\item For $T_{\rm os} > \lambda M$, the saxion can be stabilized at
  the origin solely by the effective mass term.  The condition is
  nothing but $m_{\rm s}^2 < \lambda n_Q^{s}/\phi_c$.  This condition
  is rewritten as
  \begin{equation}
    T_{\rm R} > m_{\rm s} 
    \left( \frac{\phi_i}{\lambda^2 \alpha M_{\rm pl}} \right)^{1/2}
    \sim 10^3\,\mbox{GeV} \(\frac{1}{\alpha} \)^{1/2}
    \( \frac{m_{\rm s}/\lambda}{1\,\mbox{TeV}} \)
    \( \frac{\phi_i}{10^{18}\,\mbox{GeV}} \)^{1/2}.
  \end{equation}
\item For $T_{\rm os} < \lambda M$, the combination of the effective
  mass and linear term stabilizes the saxion at the origin. The
  condition is given by $m_{\rm s}^2 M^2< \lambda n_Q^{l} M$.  This is
  rewritten as
  \begin{align}
    T_{\rm R} &> \left( \frac{m_{\rm s}^2 M}{\lambda d_\lambda} \right)^{1/3}
    \left( \frac{\phi_i}{\alpha M_{\rm pl}} \right)^{1/2} \nn
    &\sim 10^4\,\mbox{GeV} 
    \(\frac{1}{d_\lambda^{1/3} \alpha^{1/2}} \)
    \( \frac{m_{\rm s}/\lambda}{1\,\mbox{TeV}} \)^{1/3}
    \( \frac{m f_a}{10^9\, \mbox{GeV}^2} \)^{1/3}
    \( \frac{\phi_i}{10^{18}\,\mbox{GeV}} \)^{1/2}.
    \label{eq:cond_a_linear}
  \end{align}
\end{itemize}
Note that these conditions are met for most parameters of our interest
if the saxion oscillates with the thermal effects 
(recall that Eq.\eqref{eq:bdry_a_b}: $\alpha T_{\rm R} > \phi_i\sqrt{m/M_{\rm pl}}$ in the case (A)).
The saxion oscillation around the effective potential soon decays because the effective dissipation rate,
$\Gamma_\phi^{\rm (dis)}$, is large: $\Gamma_\phi^{\rm (dis)} \sim \lambda^2 \alpha T$.
Eventually, the saxion is trapped at the origin.

%%%%%%%%%%%%%%%%%%%%%%%%%%%%%%%%%%%%%%%%%%%%%%%%%
\subsubsection{Case (B): Quasi-thermal trapping}
\label{sec:trap_quasith}
%%%%%%%%%%%%%%%%%%%%%%%%%%%%%%%%%%%%%%%%%%%%%%%%%

Next, we consider the case (B).  In this case, the thermal effects on
the saxion potential is not important at the beginning of oscillation.
Still, however, the situation is similar to the case (A) studied
above.  When the saxion passes through the origin, the thermal bath
creates PQ (s)quarks.  To see this, let us estimate the time scale
$\delta t$, during which the saxion passes through the region $|\phi|
\lesssim \phi_c \equiv T_{\rm os}/\lambda$.  It is estimated as
$\delta t \sim T_{\rm os}/k_*^2$ where $k_* =\sqrt{\lambda m \phi_i}$.
The number density of PQ (s)quarks produced during this time interval is given by $n_Q \sim n_{\rm th}^2 
\langle\sigma_{\rm prod} v\rangle \delta t$, and it is estimated as
\begin{equation}
	n_Q^{s} \simeq 
	{\rm min}\left[1,\frac{g^2}{\epsilon}\right] \times T_{\rm os}^3 \simeq d_\epsilon T_{\rm os}^3,
\end{equation}
where $\epsilon \equiv k_*^2/(g^2 T_{\rm os}^2)$ and $d_\epsilon
\equiv (1+\epsilon/g^2)^{-1}$.  Note that $\lambda < \epsilon < 1$ in
the case (B).  If $\epsilon > g^2$, PQ (s)quarks does not reach
thermal equilibrium during $\delta t$ and hence the annihilation of PQ
(s)quarks is neglected after the passage of $|\phi| \simeq \phi_c$.
If $\epsilon < g^2$, PQ (s)quarks are thermally populated and their
pair annihilation takes place after the saxion climbs up the potential
around $|\phi|\sim \phi_c$.  As a result, the number density of PQ
(s)quark is given by
\begin{align}
	n_Q^{l} \simeq d_\epsilon' T_{\rm os}^3,
\end{align}
where $d_\epsilon'$ encodes the model dependent numerical factor:
$d_\epsilon' \sim g^2/\epsilon$ for $\epsilon  > g^2$, and 
$d_\epsilon' \sim \lambda^2/(\epsilon \alpha^3)$ for $\epsilon < g^2$ 
and $\epsilon^2 \alpha^2 < \lambda^2 < \epsilon \alpha^3$.
The resulting effective potential of the saxion is same as $(\ref{Veff})$.

Then, the condition for saxion trapping is summarized as follows:
\begin{itemize}
	\item For $T_{\rm os} > \lambda M$, the saxion can be stabilized at the origin 
	solely by the effective mass term.
	The condition is nothing but $m_{\rm s}^2 < \lambda n_Q^s/\phi_c$.
	This condition is rewritten as
	\begin{align}
		T_{\rm R} >  \frac{m_{\rm s}^2}{\lambda^2 d_\epsilon \sqrt{m M_{\rm pl}}}
		\sim 10^{-3} \,\mbox{GeV} \(\frac{1}{d_\epsilon}\) \( \frac{m_{\rm s}/\lambda}{1\,\mbox{TeV}} \)^2
		\( \frac{m}{1\,\mbox{GeV}} \)^{-1/2}.
	\end{align}
	\item For $T_{\rm os} < \lambda M$, the combination of the effective mass and linear term stabilizes
	the saxion at the origin. The condition is given by $m_{\rm s}^2 M^2< \lambda n^l_Q M$.
	This is rewritten as
	\begin{align}
		T_{\rm R} &> 
		\left(\frac{m_{\rm s}}{\lambda d_\epsilon'}\right)^{2/3}
		\frac{(m^{1/4} f_a)^{2/3}}{\sqrt{M_{\rm pl}}} \nonumber \\
		&\sim
		10^{-1} \,\mbox{GeV}\, d_\epsilon'^{-2/3}
		\( \frac{m_{\rm s}/\lambda}{1\,\mbox{TeV}} \)^{2/3}
		\( \frac{m}{1\,\mbox{GeV}} \)^{1/6} \( \frac{f_a}{10^9\,\mbox{GeV}} \)^{2/3}
	\end{align}
\end{itemize}
for $mM_{\rm pl} > T_{\rm R}^2$.  Note that these conditions are met
for most parameters of our interest.  Similar to the case (A), the
saxion oscillation around the effective potential soon loses its
energy due to the dissipation and is trapped at the origin.

%%%%%%%%%%%%%%%%%%%%%%%%%%%%%%%%%%%%%%%%%%%%%%%%%
\subsubsection{Case (C): Non-thermal trapping}
\label{sec:trap_nonth}
%%%%%%%%%%%%%%%%%%%%%%%%%%%%%%%%%%%%%%%%%%%%%%%%%

Finally we consider the case (C) ($k_* > gT_{\rm os}$).
Even if the above thermal effects are not efficient, the saxion can be trapped at the origin
due to the explosive production of PQ squarks when the saxion passes through its origin.
Let us see this phenomenon, which was dubbed as ``non-thermal phase transition''~\cite{Kofman:1995fi}.

As discussed in Sec.~\ref{sec:NP}, PQ squarks are efficiently produced
for $k_* > gT_{\rm os}$ for each passage of $|\phi|=0$ and its number
density exponentially grows until the condition $k_*^2 \simeq m_{{\rm
    eff},Q}^{2}$ is saturated.  After the saturation, the amplitude of
$\phi$ is given by
\begin{equation}
  \tilde\phi_s \simeq c \sqrt{\frac{m\phi_i}{\lambda}}.
  % ,~~~c \equiv 1 + \frac{g^2}{\lambda^2}.
\end{equation}
[For the definition of $c$, see Eq.\ \eqref{eq:pr_finish}.]  Using
this, the number density of PQ squark is given by $n_Q \sim \lambda
\tilde\phi_s^3/c^4$.  The saxion effective potential is given by
\begin{align}
  V_{\rm eff} \simeq
  \begin{cases}
    m_{{\rm eff}}^2 |\phi|^2 & \mbox{for}~~|\phi| < \phi_{\rm NP} \\
    \lambda n_Q |\phi| & \mbox{for}~~|\phi| > \phi_{\rm NP},
  \end{cases}
\end{align}
where $\phi_{\rm NP}=k_*/\lambda \sim \tilde\phi_s/c$ and 
$m_{{\rm eff}}^2 = \lambda^2 n_Q/k_* \sim \lambda^2\tilde\phi_s^2/c^3$.
Here $k_*^2=\lambda \bar m_{{\rm eff}} \tilde \phi_s$ with
$\bar m_{{\rm eff}}^2 = \lambda n_Q /\tilde \phi \sim \lambda^2 \tilde \phi_s^2 / c^4$.
In this case, the squark pair annihilation rate is given by $\Gamma_{{\rm ann},Q} = \langle\sigma_{\rm ann} v\rangle n_{Q}$.
Since it is dominated by $\phi(t) \sim \phi_{\rm NP}$, the averaged annihilation rate is estimated as
\begin{equation}
	\bar\Gamma_{{\rm ann},Q} = \Gamma_{{\rm ann},Q}(\phi_{\rm NP}) \frac{\phi_{\rm NP}}{\tilde\phi_s}
	=\frac{g^4}{\lambda c^4}\frac{\tilde\phi_s^2}{\phi_{\rm NP}} \sim \frac{g^4}{\lambda c^3}\tilde\phi_s 
	\sim \bar m_{{\rm eff}} \frac{g^4}{\lambda^2 + g^2}.
\end{equation}
Thus the pair annihilation of squarks is not efficient in the time scale of saxion oscillation.\footnote{
	The annihilation of PQ squarks may become efficient slightly before the condition 
	$k_\ast \sim m_{{\rm eff},Q}$ is saturated, since $g$ is not small and the rate of
	parametric resonance $\mu \bar m_{\rm eff}$ is smaller than $\bar m_{\rm eff}$.
	Even if this is the case, the conclusion is almost unchanged:
	Actually, in this case, the number density of squark stops growing slightly before the saturation
	and the saxion loses its energy in each oscillation via the annihilation.
	The annihilation of PQ squarks produces background plasma.
	Thus, it is considered that the saxion oscillation soon disappears within a few Hubble time
	and that it is eventually trapped at the origin. 
	[See the following discussion in the text below Eq.~\eqref{tildephi}.]
}

Then, the condition for saxion trapping is summarized as follows:
\begin{itemize}
	\item For $\phi_{\rm NP} > M$, the saxion can be stabilized at the origin 
	solely by the mass term.
	The condition is nothing but $m_{\rm s} < m_{{\rm eff}}$.
	This condition is rewritten as
	\begin{equation}
%		 \lambda^2 \tilde \phi_s^2 
%		 > c^3 m_{\rm s}^2 ~~\leftrightarrow ~~
                 \phi_i > \frac{c m_{\rm s}^2}{\lambda m}
		 \sim 10^6\,\mbox{GeV}\, c \lambda \( \frac{m_{\rm s}/\lambda}{1\,\mbox{TeV}} \)^2 
		 \( \frac{m}{1\,\mbox{GeV}} \)^{-1}.
	\end{equation}
	\item For $\phi_{\rm NP} < M$, the combination of mass and linear terms stabilizes
	the saxion at the origin. The condition is given by $m_{\rm s}^2 < \lambda n_Q/M$.
	This is rewritten as
	\begin{align}
          % \lambda^2 \tilde \phi_s^3 > c^4 m_{\rm s}^2 M ~~\leftrightarrow ~~ 
          \phi_i > \frac{(c m_{\rm s}^2 M)^{2/3}}{\lambda^{1/3} m} 
%          \nonumber \\
          \sim 10^8\,\mbox{GeV}\,c^{2/3} \lambda^{1/3} \( \frac{m_{\rm s}/\lambda}{1\,\mbox{TeV}} \)^{2/3}
		 \( \frac{m}{1\,\mbox{GeV}} \)^{-1/3}
		 \( \frac{f_a}{10^9\,\mbox{GeV}} \)^{2/3}.
	\end{align}
\end{itemize}
These conditions are likely satisfied for parameters of our interest.

Here we study the subsequent evolution of the system after the saxion trapping.
After the saturation, the saxion loses its energy through the squark pair annihilation into radiation.\footnote{
Since the produced squarks are highly correlated and have
    non-perturbatively large occupation number, the quasi-particle
    treatment may not be justified soon after the end of preheating,
    and other processes including in-medium-bremsstrahlung and
    multi-annihilation may also
    contribute.  Here, however, we simply estimate the energy
    transportation of saxion to the radiation by the pair annihilation
    of quasi-particle squarks.
}
The energy loss in one oscillation of the saxion is estimated as
\begin{equation}
	\Delta\rho_\phi^{(Q-{\rm ann)}} = m_Q  \langle\sigma_{\rm ann} v\rangle n_{Q}^2 \Delta t
	\sim \frac{g^4}{\lambda^2} \bar m_{{\rm eff}} n_Q \sim \frac{g^4}{c^6}\tilde\phi^4,
\end{equation}
where $\Delta t = \bar m_{{\rm eff}}^{-1}$ using the fact that the energy loss is dominated around $\phi\sim \tilde \phi$.
We can define the effective dissipation rate of the saxion through
\begin{equation}
	\bar \Gamma_\phi^{(Q-{\rm ann})} = \frac{\Delta\rho_\phi^{(Q-{\rm ann)}}}{\rho_\phi} \bar m_{{\rm eff}}
	\sim \frac{g^4}{\lambda c^4}\tilde\phi.
\end{equation}
Since this is much larger than the Hubble expansion rate $(H_{\rm
  os}\sim m)$ with $\tilde \phi \sim \tilde \phi_s$, the produced PQ
squarks efficiently annihilate and correspondingly the saxion
amplitude decreases.  By solving the equation
\begin{equation}
  \dot \rho_\phi +\bar \Gamma_\phi^{(Q-{\rm ann})} \rho_\phi = 0,
\end{equation}
we find 
\begin{equation}
	\tilde \phi(t) = \left[ \tilde\phi_s^{-1}+\frac{g^4}{4\lambda c^4}(t-t_i) \right]^{-1}. \label{tildephi}
\end{equation}
Thus it is seen that the amplitude $\tilde\phi$ decays in a time scale
$(\delta t)_{\rm ann} \sim \lambda c^4/(g^4\tilde\phi_s) $, which is
much shorter than the Hubble time scale.  The amplitude follows
Eq.~(\ref{tildephi}) as long as the condition for the non-perturbative
particle production is saturated ($k_* \sim m_{{\rm eff},Q}$).  The
annihilation of PQ squarks produces background plasma, which would
give a large effective mass to the PQ squarks and terminate the
parametric resonance.  Actually, after a time scale $(\delta t)_{\rm
  ann}$, the radiation with a ``would be'' temperature $T^{\rm (w.b.)}
\sim \rho_\phi^{1/4} \sim \sqrt{m\phi_i}$ is produced.  (Note that one
has $T^{\rm (w.b.)} \sim k_\ast/\lambda^{1/2} > (g/\lambda^{1/2}) T >
T$.)  After that, the produced plasma soon thermalizes, and the
condition for the non-perturbative particle production is saturated by
the effective mass term for the PQ squark, as $k_*^2 \sim g^2 T^2$.
Similar to the above estimates, the saxion also efficiently dissipates
its energy into radiation in this case.  Therefore it is considered
that the saxion oscillation soon disappears within a few Hubble time
around the steep effective potential.  Since the temperature is much
larger than the soft mass scale $m_{\rm s}$, the saxion is eventually
trapped by thermal mass term even if the reheating temperature is
extremely low.

After all, for all the cases (A)--(C), the saxion is likely trapped for
most parameters of our interest.

%%%%%%%%%%%%%%%%%%%%%%%%%%%%%%%%%%%%%%%%%%%%%%%%%
\subsection{Saxion dynamics: large mixing case}
\label{sec:large}
%%%%%%%%%%%%%%%%%%%%%%%%%%%%%%%%%%%%%%%%%%%%%%%%%

Now let us study the case of large mixing between PQ and SM quarks, so
that PQ (s)quarks are unstable.  Below we analyze the dynamics of the
saxion in the cases (A) -- (C) separately as in the small mixing case.

%%%%%%%%%%%%%%%%%%%%%%%%%%%%%%%%%%%%%%%%%%%%%%%%%
\subsubsection{Case (A)}
%%%%%%%%%%%%%%%%%%%%%%%%%%%%%%%%%%%%%%%%%%%%%%%%%

In the case (A), the saxion begins to oscillate with thermal log
potential.  As explained in Sec.~\ref{sec:trap_th}, the time scale
required for the saxion to pass through $|\phi| \lesssim \phi_c =
T_{\rm os}/\lambda$ is estimated as $\delta t \sim 1/(\lambda\alpha
T_{\rm os})$.  The number density PQ (s)quarks produced from thermal
plasma in this time scale is given by $n_Q \sim n_{\rm th}^2
\langle\sigma_{\rm prod} v\rangle \delta t \sim \kappa^2 T_{\rm os}^3
/ \lambda$.  (In the following, we assume $\kappa\sim 1$.)  Thus PQ
(s)quarks are expected to be thermally populated.  After the passage,
the saxion may climb up the linear potential $V_{\rm eff} = \lambda
T_{\rm os}^3 |\phi|$ for $|\phi| > \phi_c$, and eventually reach its
maximum value in $\delta t_{\rm max} \sim \alpha/ (\lambda T)$.  If
the decay time is longer than this time scale, the saxion moves back
to its origin by the linear potential and consequently oscillates with
the linear potential.  Otherwise, the produced (s)quarks soon decay
and the saxion oscillates with the thermal log potential.  This
condition is given by $\kappa^2 > \lambda / \alpha$ and we
concentrate on this case unless otherwise stated.

Note that the effective dissipation rate of the saxion in the regime
of thermal log potential is given by $\Gamma_\phi^{\rm (dis)} \sim
b\alpha^2 T^3 /(\tilde\phi \phi_c)$ and hence we obtain
$\Gamma_\phi^{\rm (dis)} / \bar m_{\rm eff} \sim \lambda\alpha$.  It
means that the saxion oscillation soon dissipates its energy once it
starts to oscillate around the origin unless $\lambda$ is very small.
This is because the effective dissipation rate decrease more slowly
than the Hubble expansion rate.  As a result, the saxion is trapped at
the origin.

Thus, the condition for the saxion trapping is summarized as follows.
\begin{itemize}
\item For $T_{\rm os} > \lambda M$, the saxion can be stabilized at
  the origin solely by the thermal mass term for $m_{\rm s} < \lambda
  T_{\rm os}$.  This condition is rewritten as
  \begin{equation}
    T_{\rm R} > \frac{m_{\rm s}}{\lambda} 
    \left( \frac{\phi_i}{\alpha M_{\rm pl}} \right)^{1/2}
    \sim 10^3\,\mbox{GeV}\, 
    \( \frac{1}{\alpha^{1/2}} \)
    \( \frac{m_{\rm s}/\lambda}{1\,\mbox{TeV}} \)
    \( \frac{\phi_i}{10^{18}\,\mbox{GeV}} \)^{1/2}.
  \end{equation}
\item For $T_{\rm os} < \lambda M$, the combination of the thermal
  mass and the linear potential/the thermal log stabilizes the saxion
  at the origin. The condition is given by $m_{\rm s}< \alpha T_{\rm
    os}^2/M$.  This is rewritten as
  \begin{equation}
    T_{\rm R} > 
    % \begin{cases}
    \left( \cfrac{mf_a\phi_i}{\alpha^2 M_{\rm pl}} \right)^{1/2}
    \sim 10^6\,\mbox{GeV}\, 
    \(\frac{0.1}{\alpha} \)
    \( \cfrac{m f_a}{10^{10}\,\mbox{GeV}^2} \)^{1/2}
    \( \cfrac{\phi_i}{10^{18} \,\mbox{GeV}} \)^{1/2}.
  \end{equation}
\end{itemize}
Here we have omitted the case $\lambda/\alpha < \kappa^2$ since it
is the same as Eq.~\eqref{eq:cond_a_linear} for $d_\lambda = 1$.  

%%%%%%%%%%%%%%%%%%%%%%%%%%%%%%%%%%%%%%%%%%%%%%%%%
\subsubsection{Case (B)}
%%%%%%%%%%%%%%%%%%%%%%%%%%%%%%%%%%%%%%%%%%%%%%%%%

In the case (B), the saxion begins to oscillate with zero-temperature
mass.  As explained in Sec.~\ref{sec:trap_quasith}, the time interval
required for the saxion to pass through the region $|\phi| \lesssim
\phi_c = T_{\rm os}/\lambda$ is estimated as $\delta t \sim T_{\rm
  os}/k_*^2$.  The number density PQ (s)quarks produced from thermal
plasma in this time period is given by $n_Q \sim n_{\rm th}^2
\langle\sigma_{\rm prod} v\rangle \delta t \sim (\kappa^2\alpha T_{\rm
  os}^2/k_*^2) T_{\rm os}^3$.  By noting $(1/\lambda^2\gtrsim) \alpha
T_{\rm os}^2/k_*^2 \gtrsim 1$ in the case (B), we find that PQ
(s)quarks are expected to be thermally populated in this time
interval.  For most parameters of our interest, the linear potential
may soon decay, and hence the effective potential for the saxion near
the origin is given by thermal potential.\footnote{
  In fact, the linear potential can decay if $1 < \Gamma^Q_{\rm dec}
  \delta t$ with $\Gamma^Q_{\rm dec} \sim \kappa^2 \lambda
  |\phi|$ and $\delta t \sim m\tilde \phi / (\lambda T^3)$.
  Such a condition is likely satisfied because
  $\Gamma^Q_{\rm dec} \delta t \sim (\kappa^2/ \lambda)
  (m \tilde \phi / T^2) > \kappa^2 \alpha /\lambda$.}

The effective dissipation rate of the saxion in the regime is
estimated as $\Gamma_\phi^{\rm (dis)} \sim \lambda \alpha T^2/\tilde
  \phi$ and hence we obtain $\Gamma_\phi^{\rm (dis)} / m \sim
\lambda \alpha T_{\rm os}^2/(m\tilde\phi)$. 
It leads to $\lambda^2 < \Gamma_\phi^{\rm (dis)} / m <1$.  
Thus the saxion oscillation soon dissipates its energy
unless $\lambda$ is very small since $\Gamma_\phi^{\rm (dis)}$
decreases more slowly than the Hubble rate.

In summary, the condition for the saxion trapping is as follows:
\begin{itemize}
\item For $T_{\rm os} > \lambda M$, the saxion can be stabilized at the origin 
  solely by the effective mass term for $m_{\rm s} < \lambda T_{\rm os}$.
  This condition is rewritten as
  \begin{equation}
    T_{\rm R} > \frac{1}{\lambda^2} \frac{m_{\rm s}^2}{\sqrt{mM_{\rm pl}}}
    \sim 10^{-3}\,\mbox{GeV} 
    \( \frac{m_{\rm s} / \lambda }{1\, \mbox{TeV}} \)^2
    \( \frac{m}{1\,\mbox{GeV}} \)^{-1/2}.
  \end{equation}
\item For $T_{\rm os} < \lambda M$, the combination of the effective
  mass and thermal log stabilizes the saxion at the origin. The
  condition is given by $m_{\rm s}< \alpha T_{\rm os}^2/M$.  This is
  rewritten as
  \begin{equation}
    T_{\rm R} > \frac{f_a}{\alpha}\left(\frac{m}{M_{\rm pl}} \right)^{1/2}
    \sim 10\,\mbox{GeV} \( \frac{0.1}{\alpha} \) 
    \( \frac{f_a}{10^9\,\mbox{GeV}} \)
    \( \frac{m}{1\,\mbox{GeV}} \)^{1/2}.
  \end{equation}
\end{itemize}
%%

%%%%%%%%%%%%%%%%%%%%%%%%%%%%%%%%%%%%%%%%%%%%%%%%%
\subsubsection{Case (C)}
%%%%%%%%%%%%%%%%%%%%%%%%%%%%%%%%%%%%%%%%%%%%%%%%%

In the case (C), the non-perturbative production occurs at the
crossing of $\phi\sim 0$.  As explained in Sec.~\ref{sec:NP}, the
number density of PQ (s)quarks produced in this way is given by $n_Q
\sim k_*^3/(4\pi^3)$.  The produced PQ (s)quarks decay at
$|\phi(t_{\rm dec})|\sim[m\tilde\phi/\lambda\kappa^2]^{1/2}$.
From this, we can estimate the effective dissipation rate of the
saxion as~\cite{Mukaida:2012qn}
\begin{equation}
  \Gamma_\phi^{\rm (dis)} \simeq N_{\rm d.o.f.}
  \times \frac{\lambda^2 m}{2\pi^4\kappa}.
\end{equation}
Due to this effect, thermal plasma with temperature $T\sim \sqrt{\lambda m \phi_i/g^{1/2}}$ is produced within one oscillations of the saxion.
This process continues until the condition for the non-perturbative production is violated.
Actually the condition is soon violated after a few Hubble time after the oscillation, that is, 
$g T_{\rm NP} \sim k_\ast$\footnote{
	The interaction given in Eq.~\eqref{eq:mix} also affects the thermal mass
	of $Q$.  Even with such a contribution, the present discussion is unchanged
	because we are considering the case of $g\sim \kappa\sim 1$.
} [See Eq.~\eqref{NP:begin}].
Then, thermal effects tend to stabilize the saxion at the origin, similar to the case (A) and (B).

By simply assuming that the temperature of the plasma at the end of non-perturbative particle production 
is given by $T_{\rm NP} \sim \sqrt{\lambda m\phi_i}/g$, the condition for the saxion trapping is summarized as follows.
\begin{itemize}
	\item For $T_{\rm NP} > \lambda M$, the saxion can be stabilized at the origin 
	solely by the effective mass term for 
	$m_{\rm s}/\lambda <  T_{\rm NP} \sim \sqrt{\lambda m\phi_i}/g$.
	\item For $T_{\rm NP} < \lambda M$, the combination of the effective mass and thermal log stabilizes
	the saxion at the origin. The condition is given by $m_{\rm s}< \alpha T_{\rm NP}^2/M$.
	This condition is rewritten as $4\pi f_a < \lambda \phi_i$.
\end{itemize}
These conditions are likely satisfied in the most parameters of our interest.
Then, as in the case of (A) and (B), the saxion soon dissipates its energy unless $\lambda$
is quite small, and sits around its origin.
Once the saxion is trapped at the origin, it potentially causes thermal inflation, as studied in the next section.

%%%%%%%%%%%%%%%%%%%%%%%%%%%%%%%%%%%%%%%%%%%%%%%%%
\subsubsection{Comment on the case of non-trapped saxion}
%%%%%%%%%%%%%%%%%%%%%%%%%%%%%%%%%%%%%%%%%%%%%%%%%

As studied above, it is possible that the saxion oscillates with an amplitude much larger than the PQ scale
without trapping at the origin in the limit of small $\lambda$ and $T_{\rm R}$.
Then it may be non-trivial in which minima the saxion relaxes and whether axionic domain walls forms or not.
We briefly see what happens in this case.

Let us suppose that the saxion oscillates in the real axis in $\phi$ with an amplitude $\tilde\phi$
and define the axion as $a \equiv \sqrt{2}{\rm Im \phi}$. 
First note that the mass of $a$ changes from $+m^2$ to $-m^2$ around $|\phi| \simeq f_a$ and further it develops to
$-m_{\rm s}^2$ at $|\phi| \lesssim M$.
In one oscillation of the saxion, the fluctuation along the axion direction develops due to the tachyonic instability
as $\delta a / a \sim f_a / \tilde \phi$ at $|\phi| \lesssim f_a$.
Therefore, the initial axion fluctuation sourced by inflationary quantum fluctuations significantly develops at $\tilde\phi \sim f_a$
and the motion of $\phi$ is expected to become chaotic in the complex plane.
This may lead to the formation of axionic domain walls, even in the case of no explicit symmetry restoration.

A further comment is in order.  It was pointed out that the parametric
resonant decay of the saxion into the axion leads to the nonthermal
phase transition \cite{Kasuya:1996ns,Kasuya:1997ha,Tkachev:1998dc}.
Note that these results were based on the potential $V \propto
(|\phi|^2-f_a^2)^2$.  In our setup, the interaction term between the
saxion and axion exists only in the logarithmic term and the
parametric resonant enhancement of the axion would be much less
efficient.  Actually, the condition for the broad resonance is only
marginally satisfied in our case.  In order to make a definite
conclusion whether the parametric resonance of the axion is efficient
or not, we may need lattice calculations, which is beyond the scope of
our analysis.  However, regardless of the efficiency of the parametric
resonance, we expect the formation of axionic domain walls as
explained above.

%%%%%%%%%%%%%%%%%%%%%%%%%%%%%%%%%%%%%%%%%%%%%%%%%
\section{Saxion Dynamics after Phase Transition}
\label{sec:PT}
%%%%%%%%%%%%%%%%%%%%%%%%%%%%%%%%%%%%%%%%%%%%%%%%%

%%%%%%%%%%%%%%%%%%%%%%%%%%%%%%%%%%%%%%%%%%%%%%%%%
\subsection{Condition for thermal inflation}
%%%%%%%%%%%%%%%%%%%%%%%%%%%%%%%%%%%%%%%%%%%%%%%%%

In the previous section, we saw that the saxion is trapped at the origin $\phi=0$ for almost all the cases
even if the initial position of the saxion is displaced far from the origin.
Now let us examine whether it causes thermal inflation or not.

After the trapping, the saxion begins to roll down the potential toward the minimum $|\phi|=f_a$ at the temperature
$T=T_{\rm PT} \simeq m_{\rm s}/\lambda$.
The condition for thermal inflation by the saxion to occur is $\rho_{\rm rad, inf}(T=T_{\rm PT}) < m_{\rm s}^2 M^2 (\simeq m^2 f_a^2)$
where $\rho_{\rm rad, inf}$ denotes the energy density of radiation or the inflaton oscillation, whichever is dominant.

First, let us consider the low-reheating temperature case: $T_{\rm PT} > T_{\rm R}$, so that the condition is given by
$3H_{\rm PT}^2M_{\rm pl}^2 < m_{\rm s}^2 M^2$.
Then, the radiation component in this case may contain two contributions: 
one from the dilute plasma due to the inflaton decay and the other from the dissipation of the saxion.
The former (latter) is denoted by $\rho_r^{\rm (inf)}$ $(\rho_r^{\rm (\phi)})$.
If $\rho_r^{\rm (inf)}$ dominates the radiation energy density at $T=T_{\rm PT}$,
it is found that thermal inflation takes place if
\begin{equation}
	T_{\rm R} > \frac{T_{\rm PT}^2}{\sqrt{mf_a}} \sim 10\,{\rm GeV} 
	\left( \frac{m_{\rm s} / \lambda}{1\,{\rm TeV}} \right)^2 \left( \frac{1\,{\rm GeV}}{m} \right)^{1/2}\left( \frac{10^{10}\,{\rm GeV}}{f_a} \right)^{1/2}.
\end{equation}
On the other hand, if $\rho_r^{\rm (\phi)}$ dominates the radiation energy density at $T=T_{\rm PT}$,
by noting that $\rho_r^{\rm (\phi)} (T_{\rm PT}) \simeq m^2 \phi_i^2 (H_{\rm PT}/m)^{8/3}$, the condition is given by
\begin{equation}
	f_a > \frac{M_{\rm pl}(m_{\rm s}/\lambda)^{3/2}}{(m\phi_i)^{3/4}}
	\sim 10^9\,{\rm GeV}
	\left( \frac{m_{\rm s} / \lambda}{1\,{\rm TeV}} \right)^{3/2}
	\left( \frac{1\,{\rm GeV}}{m} \right)^{3/4}
	\left( \frac{M_{\rm pl}}{\phi_i} \right)^{3/4}.
\end{equation}

Second, let us consider the radiation dominant case: $T_{\rm PT} < T_{\rm R}$.
Then the condition is given by $T_{\rm PT}^4 < m_{\rm s}^2 M^2$. This condition is rewritten as
\begin{align}
	f_a > \frac{1}{m} \( \frac{m_{\rm s}}{\lambda} \)^2
	\sim 10^6\, \mbox{GeV} \( \frac{1\, \mbox{GeV}}{m} \) \( \frac{m_{\rm s} / \lambda}{1\,\mbox{TeV}} \)^2.
	\label{TIcond}
\end{align}

If either of these conditions are met, thermal inflation is caused by the saxion.
Actually, unless the reheating temperature is very low or the mass $m$ is small, thermal inflation likely takes place.
Otherwise, the saxion exits the thermal trapping potential before it begins to dominate the Universe
and the cosmological problems associated with the axion overproduction from the saxion decay,
as studied in detail in the next subsection, is alleviated.
Even in such a case, however, the saxion may eventually dominate the Universe before it decays.
In the following, we mainly consider the case of thermal inflation where the problem is most prominent.

%%%%%%%%%%%%%%%%%%%%%%%%%%%%%%%%%%%%%%%%%%%%%%%%%
\subsection{Saxion dynamics after phase transition}
%%%%%%%%%%%%%%%%%%%%%%%%%%%%%%%%%%%%%%%%%%%%%%%%%

Now let us see the cosmological evolution after the saxion begins to roll down from the high-temperature minimum $\phi=0$
toward the true minimum $|\phi| = f_a$.
For clarity, we assume $T_{\rm R} > T_{\rm PT}$ hereafter;
this is the case in the most of parameters of our interest.
We divide the saxion dynamics into two stages.
At the first stage, the lower edge of the saxion oscillation field value is smaller than $M$.
At the second stage, the lower edge of the saxion oscillation field value is larger than $M$.

%%%%%%%%%%%%%%%%%%%%%%%%%%%%%%%%%%%%%%%%%%%%%%%%%
\subsubsection{First stage}
%%%%%%%%%%%%%%%%%%%%%%%%%%%%%%%%%%%%%%%%%%%%%%%%%

We consider the dynamics of the saxion after it exits the thermal trapping potential at $T=T_{\rm PT} \sim m_{\rm s}/\lambda$.
Let us define the time $t$ so that $t=0$ [$T(t=0) = T_{\rm PT}$]
as the initial condition and reaches $|\phi| = M (f_a)$ at $t=t_M (t_f)$.
Then, we can evaluate the dissipative effect on the saxion during its one oscillation around the minimum, 
using the fact that the background temperature is (almost) unchanged for the time scale of one oscillation.

\begin{itemize}
\item $t< t_{M}$\\
  The evolution of the saxion is given by $|\phi (t)| = \phi_0
  \exp(m_{\rm s}t)$ where $\phi_0$ is expected to be of order $T$ (although
  its precise value is not important).  Thus $t_M$ is given by $t_M =
  m_{\rm s}^{-1} \log(M/\phi_0)$.  The energy loss of saxion due to
  the dissipation effect when it passes through the field value around
  $\phi$ is given by\footnote
  {The temperature $T \sim m_{\rm s}/\lambda$ is slightly larger than
    the mass scale $m_{\rm s}$, hence the use of dissipation rate in
    thermal background is marginally justified.}
  \begin{align}
    \left[\Delta \rho_\phi^{\rm (dis)} \right]_{0 < t < t_{\rm M}} 
    \sim \int_0^{t_{\rm M}} d t\,
    \frac{1}{2}\dot\phi^2 \times \frac{b\alpha^2T^3}{\phi^2} 
    \sim b \alpha^2 T^3 m_{\rm s} \log\left( \frac{M}{\phi_0} \right), 
  \end{align}
  where $b\alpha^2 = 9\alpha^2/(128\pi^2\ln\alpha^{-1})\sim 3\times
  10^{-5}$ \cite{Laine:2010cq}.
\item $t_{M} < t < t_f$\\
  After the passage of $|\phi|=M$, the saxion ``speed'' $\dot \phi
  (\sim m_{\rm s}M)$ remains nearly constant until it reaches
  $|\phi|=f_a$.  Therefore, we have $|\phi(t)| = M[1+m_{\rm
    s}(t-t_M)]$ at this regime.  We can evaluate the energy loss in a
  similar way as
  \begin{align}
    \left[\Delta \rho_\phi^{\rm (dis)} \right]_{t_M < t < t_f} 
    \sim \int_{t_{\rm M}}^{t_{\rm F}}dt\,
    \frac{1}{2}\dot\phi^2 \times \frac{b\alpha^2T^3}{\phi^2}
    \sim b \alpha^2 T^3 m_{\rm s}
  \end{align}
\item $t_f < t$\\
  Finally, the saxion climbs up the potential at $|\phi|>f_a$ at
  $t=t_f \sim f_a/(m_{\rm s}M) \sim 1/m$.  The dynamics of saxion at
  $|\phi|>f_a$ is described by the harmonic oscillation form with a
  frequency $m$.  Parametrically, we have $|\phi| \sim f_a$ and
  $|\dot\phi| \sim mf_a (\sim m_{\rm s}M)$.  Thus we obtain
  \begin{align}
    \left[\Delta \rho_\phi^{\rm (dis)} \right]_{t > t_f} 
    \sim \int_{t_{\rm F}} dt\, \frac{1}{2}\dot\phi^2 \times 
    \frac{b\alpha^2T^3}{\phi^2}
    \sim b \alpha^2 T^3 m.
  \end{align}
  This is smaller than the energy losses in the regime of $t<t_f$ for
  $m < m_{\rm s}$.
\end{itemize}

From the above estimates, it is found that the saxion energy loss due
to the dissipation effect in its one oscillation is given by
\begin{equation}
	\left[\Delta \rho_\phi^{\rm (dis)} \right]_{\Delta t = m^{-1}} \sim b \alpha^2 T^3 m_{\rm s} \log\left( \frac{M}{\phi_0} \right).
\end{equation}
We define the effective dissipation rate averaged over the saxion oscillation period:
\begin{equation}
	\Gamma_\phi^{\rm (dis,1)} \equiv 
	\frac{m}{\rho_\phi}\left[\Delta \rho_\phi^{\rm (dis)} \right]_{\Delta t = m^{-1}} =
	 \frac{b'\alpha^2 T^3}{ f_a^2} \frac{m_{\rm s}}{m},
\end{equation}
where $b' \equiv b \log(M/\phi_0)$.

Now let us show the evolution of the radiation and saxion energy density at a few Hubble time after the thermal inflation.
The radiation energy density evolves as
\begin{equation}
	\dot\rho_{\rm rad} + 4H\rho_{\rm rad} =\Gamma_\phi^{\rm (dis,1)}(T) \rho_\phi. \label{drho_rad}
\end{equation}
If $\Gamma_\phi^{\rm (dis,1)} \gg H_{\rm PT}$, 
the significant fraction of the saxion energy density soon goes into radiation
and the resulting radiation temperature becomes $T=T_{\rm TI} \sim \sqrt{mf_a}$.
In the opposite limit, after a few Hubble time, the radiation temperature becomes
\begin{equation}
	T = \max\left[ 1, f  \right] T_{\rm PT} \equiv F T_{\rm PT},
\end{equation}
where
\begin{align}
	f \equiv \left( b' \alpha^2 \lambda \frac{M_{\rm pl}}{f_a}\right)
	 \sim 10^9 \( b' \alpha^2 \lambda \) \( \frac{10^9\, \mbox{GeV}}{f_a} \),
\end{align}
by noting that $\dot T \sim {\rm const}$ from (\ref{drho_rad}).
Therefore, we obtain
\begin{align}
	T = \min\left[ F T_{\rm PT},  T_{\rm TI}  \right],
\end{align}
at the end of first stage after thermal inflation.

By comparing the dissipation rate with the Hubble rate at the end of thermal inflation, we obtain
\begin{equation}
	\frac{\Gamma_\phi^{\rm (dis,1)}}{H_{\rm PT}} \sim 
	F^3 b' \alpha^2 \lambda \frac{ m_{\rm s}^4 M_{\rm pl}}{\lambda^4 m^2 f_a^3}
	\sim 10^2 \times (10 F^3 b' \alpha^2\lambda) \left( \frac{m_{\rm s}/\lambda}{1\,{\rm TeV}} \right)^4
	\left( \frac{1\,{\rm GeV}}{m} \right)^2
	\left( \frac{10^{9}\,{\rm GeV}}{f_a} \right)^3.
	\label{Geff/H}
\end{equation}
The ratio (\ref{Geff/H}) can be much larger than unity, meaning that
the saxion can dissipate most of its energy within one Hubble time
after the thermal inflation ends and the radiation with a temperature
$T=T_{\rm TI}$ is produced.  Otherwise, the saxion coherent
oscillation soon dominates the Universe and it decays into the axion
pair, as we will see below, leading to unsuccessful reheating.
Therefore we must have $\Gamma_\phi^{\rm (dis,1)}/H_{\rm PT} \gg 1$.

However, even if $\Gamma^{\rm (dis,1)}$ is initially much larger than
$H_{\rm PT}$, all the energy of the saxion oscillation is not
dissipated away; this is because the dissipation rate is significantly
suppressed once the smallest value of $\phi$ during one oscillation
becomes larger than $M$.  Thus, the coherent oscillation around
$|\phi|=f_a$, whose energy density is the same order of the initial
total saxion energy density $(\rho_\phi\sim m_{\rm s}^2M^2\sim
m^2f_a^2)$, still remains after such a dissipation.  Then we need to
consider the second stage of saxion reheating.

%%%%%%%%%%%%%%%%%%%%%%%%%%%%%%%%%%%%%%%%%%%%%%%%%
\subsubsection{Second stage}
\label{sec:second}
%%%%%%%%%%%%%%%%%%%%%%%%%%%%%%%%%%%%%%%%%%%%%%%%%

At the second stage, the saxion oscillates around the minimum $|\phi| =f_a$.
Here we assume that the condition $\Gamma_\phi^{\rm (dis,1)} \gg H_{\rm PT}$ is satisfied and hence $T=T_{\rm TI}$.
The thermal dissipation rate at this epoch is given by
\begin{equation}
	 \Gamma_\phi^{\rm (dis,2)} = \frac{b \alpha^2 T^3}{ f_a^2}.
\end{equation}
By comparing it with the Hubble rate at the end of thermal inflation, we obtain
\begin{equation}
	\frac{\Gamma_\phi^{\rm (dis,2)}}{H_{\rm PT}} \sim \frac{b\alpha^2 m^{1/2}M_{\rm pl}}{f_a^{3/2}}
	\sim 10^3 b \( \frac{\alpha}{0.1} \)^2
	\left( \frac{m}{1\,{\rm GeV}} \right)^{1/2}
	\left( \frac{10^{9}\,{\rm GeV}}{f_a} \right)^{3/2}.
	\label{Geff/H2}
\end{equation}
Here we have substituted $T=T_{\rm TI}$.

At the second stage, as well as the dissipation process due to the thermal bath, there is another important process which reduces the energy density of the saxion oscillation.  In the present analysis, we consider the case where the saxion dominantly decays into the axion pair.  Then, the decay rate of the saxion is estimated as
\begin{equation}
	\Gamma_{\phi \to 2a} = \frac{1}{64\pi}\frac{m^3}{f_a^2}.
\end{equation}
The decay must happen before the saxion again comes to dominate the Universe for the successful thermal history.
Note that the ratio between $\Gamma_{\phi \to 2a}$ and $\Gamma_\phi^{\rm (dis,2)}$ is given by
\begin{equation}
	\frac{\Gamma_\phi^{\rm (dis,2)}(T_{\rm TI})}{\Gamma_{\phi \to 2a}} 
	= 64\pi b\alpha^2\left(\frac{f_a}{m}\right)^{3/2},
\end{equation}
and it is much larger than unity, hence the decay into axion is
negligible just at the end of thermal inflation.  However, as the
temperature decreases, the decay rate into the axion pair becomes
dominant.  In particular, even if $\Gamma^{\rm (dis,2)}_\phi$ becomes
larger than the expansion rate just after the thermal inflation, the
saxion coherent oscillation cannot be fully dissipated away because
the dissipation rate becomes suppressed with the decrease of the
cosmic temperature.  (See Appendix.)  Thus, the decay of the saxion
may overproduce the axion.  If the saxion still does not dominate the
Universe when the perturbative decay into the axion becomes efficient
($H \sim \Gamma_{\phi \to 2a}$), the axion overproduction is avoided.

\begin{figure}
\begin{center}
%\vskip -1.cm
\includegraphics[scale=1.2]{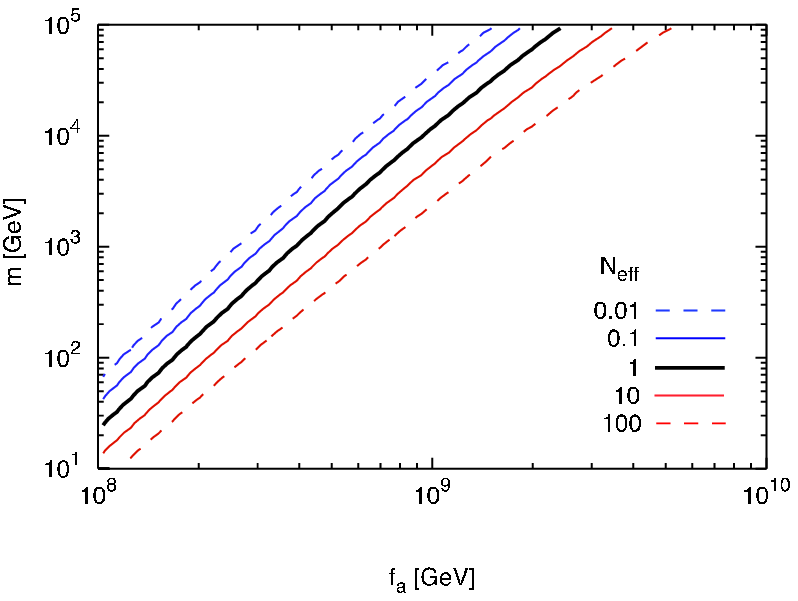}
\includegraphics[scale=1.2]{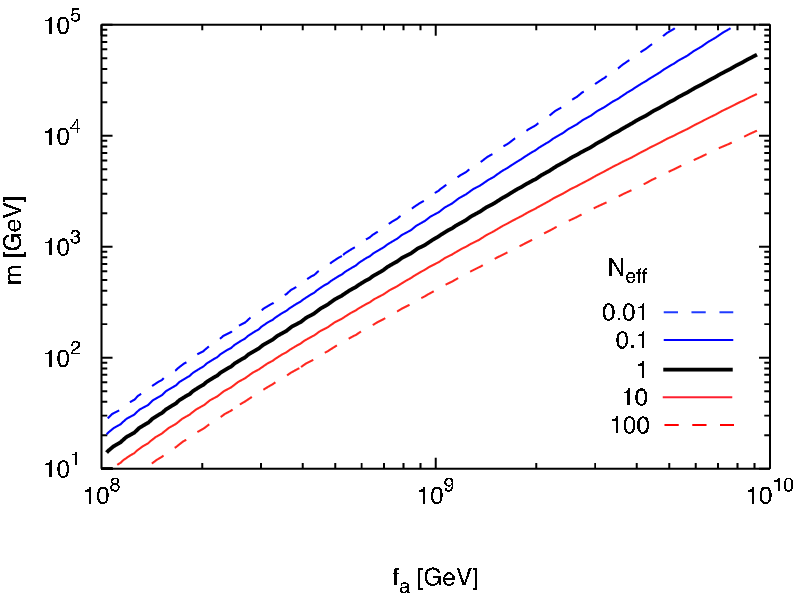}
\vskip -0.1cm
\caption{
Contours of $N_{\rm eff}$ on $(f_a, m)$ plane.
We have taken $f_a / M = 100$ (hence $m_{\rm s} \sim 100 m$) 
and $\lambda = 1$ (top) and $\lambda = 0.05$ (bottom).
}
\label{fig:Neff}
\end{center}
\end{figure}

\begin{figure}
\begin{center}
%\vskip -1.cm
\includegraphics[scale=1.0]{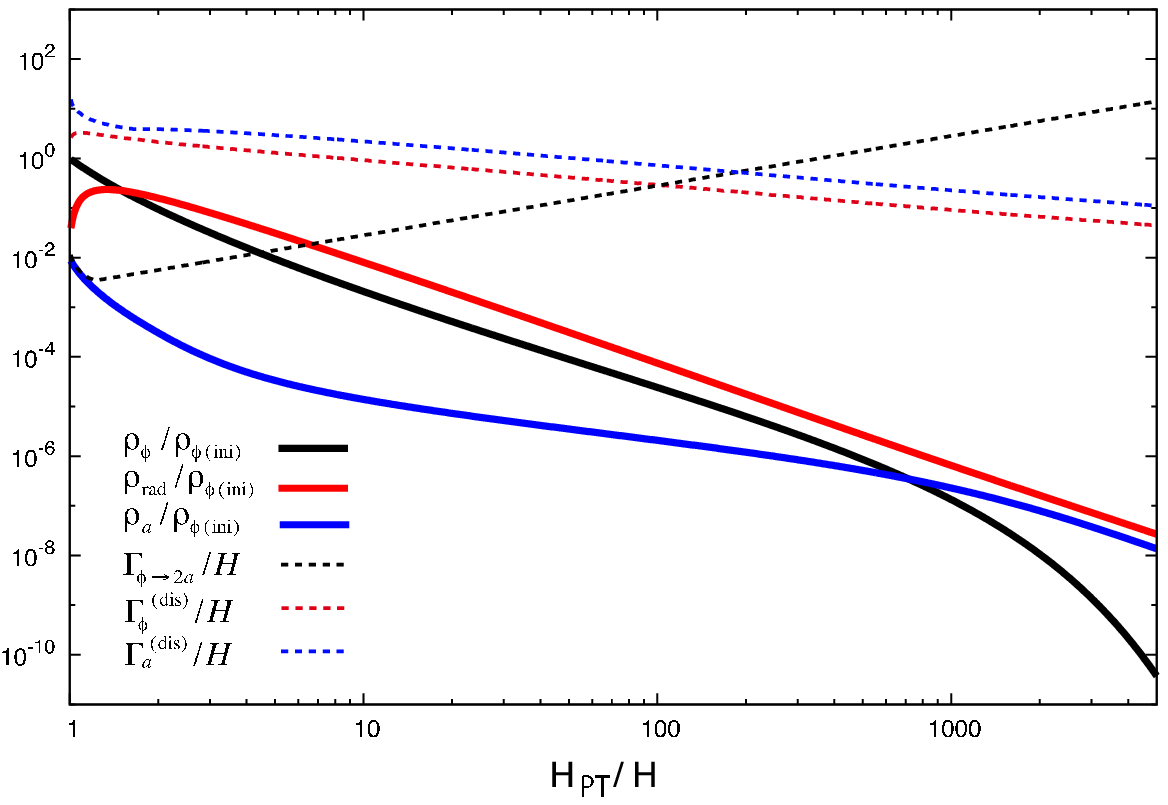}
\vskip -0.1cm
\caption{
Time evolution of various quantities as a function of $H_{\rm PT}/H$
: $\rho_{\phi}$, $\rho_{\rm rad}$, $\rho_{a}$
normalized by the initial saxion energy density $\rho_{\phi {\rm (ini)}}$,
and $\Gamma_{\phi\to 2a}$, $\Gamma_\phi^{\rm (dis)}$, $\Gamma_a^{\rm (dis)}$ normalized by $H$.
Parameters are chosen as $\lambda=1$, $m=5$\,TeV, $f_a=10^9$\,GeV, $M=10^{-2}f_a$.
}
\label{fig:time}
\end{center}
\end{figure}

In order to estimate the present energy density of axion, we have performed numerical calculation.
We have solved a set of equations
\begin{align}
	&\dot\rho_\phi + (3H + \Gamma_\phi^{\rm (dis)} + \Gamma_{\phi\to 2a})\rho_\phi= 0, \\
	%&\dot\rho_\phi + (3H + \Gamma_\phi^{\rm (dis)} + \Gamma_{\phi\to 2a})\rho_\phi = 0, \\
	&\dot\rho_{\rm rad} + 4H\rho_{\rm rad} = \Gamma_\phi^{\rm (dis)}\rho_\phi + \Gamma_a^{\rm (dis)}\rho_a, \\
	&\dot\rho_a + 4H\rho_a = \Gamma_{\phi\to 2a}\rho_\phi - \Gamma_a^{\rm (dis)}\rho_a,\\
	&3H^2 M_{\rm pl}^2 = \rho_\phi + \rho_{\rm rad} + \rho_a.
\end{align}
with initial conditions $T=T_{\rm PT}$, $\rho_a = 0$ and $\rho_\phi = m^2 f_a^2$,
where $\Gamma_a^{\rm (dis)}$ denotes the axion dissipation rate, which is roughly same as $\Gamma_\phi^{\rm (dis)}$.
Then we have calculated the extra effective number of neutrino species $\Delta N_{\rm eff}$ by
\begin{equation}
	\Delta N_{\rm eff} = \frac{43}{7}\left(\frac{10.75}{g_{*s}}\right)^{1/3}\left(\frac{\rho_a}{\rho_{\rm rad}}\right),
\end{equation}
where $g_{*s}$ is the effective number of massless degrees of freedom.
The results are plotted in Figs.~\ref{fig:Neff}.  In this plot we have
taken $f_a / M = 100$ (hence $m_{\rm s}\sim 100 m$) and $\lambda = 1$
(top) and $\lambda = 0.05$ (bottom), and $g_{*s}=228.75$.  The recent
Planck result~\cite{Ade:2013uln} excludes the parameter regions with
$\Delta N_{\rm eff} \gtrsim 1$.  It is seen that parameters with
$m\gtrsim$ a few TeV and $f_a \gtrsim 10^9$\,GeV is viable.  ($f_a
\gtrsim 4\times 10^8$\,GeV is required from astrophysical
arguments~\cite{Kawasaki:2013ae}.)  Therefore, heavy SUSY scenario
with $m \gg O(1)$\,TeV is favored.\footnote{
	Thermal inflation occurs for most of the parameter regions in the top panel of Fig.~\ref{fig:Neff}.
	In the bottom panel, thermal inflation does not take place in the about upper half of the parameter region
	(see Eq.~(\ref{TIcond})). 
}
We have checked that the figure looks quite similar for $f_a/M=1$.
To see more detail, we have plotted time evolution of various quantities as a function of $H_{\rm
  PT}/H$ in Fig.~\ref{fig:time}.  Parameters are chosen as
$\lambda=1$, $m=5$\,TeV, $f_a=10^9$\,GeV, $M=10^{-2}f_a$ with which we
obtain $\Delta N_{\rm eff} \sim 1$.  Since the dissipation rate
$\Gamma_\phi^{\rm (dis)}$ is larger than the Hubble rate at early
time, a significant amount of the saxion energy goes to the radiation.
The produced axions are also thermalized since $\Gamma_a^{\rm (dis)}$
is also large.  Although gradually $\Gamma_\phi^{\rm (dis)} / H$
decreases, the perturbative decay $\Gamma_{\phi\to 2a}$ becomes
efficient before the saxion dominates the Universe and hence the
overproduction of axion can be avoided.

Note that in a case with large dissipation rate $\Gamma_\phi^{\rm
  (dis)} \gg H$, axinos and saxions will be thermally populated.  This
does not cause cosmological problems as long as the axino is heavy
enough to decay well before BBN and the produced LSPs have relatively
large annihilation cross section \cite{Baer:2011hx, Baer:2011uz,
  Moroi:2013sla}.  All these necessary properties are consistent with
heavy SUSY scenario, such as the pure
gravity-mediation~\cite{Ibe:2011aa}, which leads to 125\,GeV
Higgs~\cite{Aad:2012tfa}.

%%%%%%%%%%%%%%%
%%%%%%%%%%%%%%%

%%%%%%%%%%%%%%%
%%%%%%%%%%%%%%%

%%%%%%%%%%%%%%%%%%%%%%%%%%%%%%%%%%%%
\section{Conclusions and Discussion}
\label{sec:conc}
%%%%%%%%%%%%%%%%%%%%%%%%%%%%%%%%%%%%

In this paper, we have investigated the dynamics of a scalar field in
thermal environment with a symmetry breaking potential.  In
particular, we have performed a detailed study of the trapping of
scalar field at an enhanced symmetry point.

Throughout this paper, we have focused on the case where the scalar
field interacts with fermion and boson which are charged under SM
gauge group and the initial amplitude of scalar field is far from the
minimum of the potential.  We considered the trapping dynamics of
scalar field with the interaction of SM plasma taken into
account. Although we have studied the SUSY PQ model in detail to
make our discussion concrete, we emphasize that our results are rather
general and that they are applicable to other symmetry breaking
dynamics than the saxion.  Since the trapping occurs due to the
particle production, there are two possible sources: the production
from the background thermal environment or the non-perturbative
production from saxion itself.  It is also notable that the higher
order correction to thermal potential becomes important, so-called
thermal logarithmic potential.  We have taken into account all these
effects carefully and found that the dynamics of scalar field
complicatedly depends on the interactions of produced particles to the
SM particles.  However, even if the saxion has a large initial
amplitude, it is found that the saxion is likely trapped at its origin
once and often leads to the thermal inflation for the most parameters
of our interest.

We have also studied the dynamics of saxion after the phase
transition. It is noticeable that, even if the saxion once dominates
the Universe and the thermal inflation occurs, the saxion can
successfully dissipate its energy by the interaction with the thermal
bath.  In such a case, we can avoid the overproduction of the axion by
the decay of the saxion field even if the dominant decay mode of the
free saxion is into axion pair. To verify this statement, we have
performed numerical calculation, and it is shown that the axion
overproduction does not occur in the parameter region that are
consistent with the high-scale SUSY scenario.

%%%%%%%%%%%%%%%%%%%%%%%%%%%%%%%%%%%%
\section*{Acknowledgment}
%%%%%%%%%%%%%%%%%%%%%%%%%%%%%%%%%%%%

This work is supported by Grant-in-Aid for Scientific
research from the Ministry of Education, Science, Sports, and Culture
(MEXT), Japan, No.\ 22244021 (T.M.), No.\ 22540263 (T.M.), No.\
23104001 (T.M.), No.\ 21111006 (K.N.), and No.\ 22244030 (K.N.).
The work of K.M. is supported in part by JSPS Research Fellowships
for Young Scientists.

%%%%%%%%%%%%%%%%%%%%%%%%%%%%%%%%%%%%
\appendix
%%%%%%%%%%%%%%%%%%%%%%%%%%%%%%%%%%%%

%%%%%%%%%%%%%%%%%%%%%%%%%%%%%%%%%%%%
\section*{Appendix}
%%%%%%%%%%%%%%%%%%%%%%%%%%%%%%%%%%%%

%%%%%%%%%%%%%%%%%%%%%%%%%%%%%%%%%%%%
\section{Time Evolution of $\rho_\phi$}
%%%%%%%%%%%%%%%%%%%%%%%%%%%%%%%%%%%%

Let us consider the evolution of the $\rho_\phi$ in the presence of time-dependent dissipation effect $\Gamma (t)$:
\begin{equation}
	\dot \rho_\phi(t) + \Gamma(t) \rho_\phi (t) = 0.
\end{equation}
Generally, $\Gamma(t)$ is parametrized as $\Gamma(t) = \Gamma_i (t_i / t)^n$ where $t_i$ is an initial time.
In the case of standard perturbative decay, $n=0$.
The case of $n=1$ includes the Hubble friction: $\Gamma(t) = \Gamma_i t_i / t$ with $\Gamma_i t_i =1/2 (2/3)$
corresponding to the RD (MD) Universe.
Thermal dissipation effect may correspond to $n > 1$.
This equation can be easily integrated to obtain
\begin{equation}
	\frac{\rho_\phi(t)}{\rho_\phi(t_i)} = \begin{cases}
		\displaystyle \left( \frac{t_i}{t} \right)^{\Gamma_i t_i} &~~{\rm for}~~n=1,\\
		\displaystyle \exp\left[ \frac{\Gamma_i t_i}{n-1}\left\{ \left(\frac{t_i}{t}\right)^{n-1}-1\right\} \right] &~~{\rm for}~~n\neq 1.
	\end{cases}
\end{equation}
From this expression it is easily checked that $\rho_\phi$ exponentially decays with time for $n<1$.
For $n>1$, the energy density exponentially decreases at $t \gtrsim {\rm (a~few)}\times t_i$ if $\Gamma_i t_i \gg 1$.
Therefore, if $\Gamma_i t_i \gg 1$, $\phi$ loses most of its energy within a few Hubble time.
Eventually $\rho_\phi$ approaches to the asymptotic value
\begin{equation}
	\rho_\phi(t) \to \rho_\phi(t_i)  \exp\left[ -\frac{\Gamma_i t_i}{n-1}\right],
\end{equation}
at $t > t_f$ where $t_f$ is defined by $t_f \Gamma(t_f) = 1$.

%%%%%%%%%%%%%%%%%%%%%%%%%%%%%%%%%%%%%%%%%%%%

%%%%%%%%%%%%%%%%%%%%%%%%%%%%%%%%%%%%%%%%%%%%

\end{document}